\documentclass[12pt]{article}

\usepackage{scicite}

\usepackage{graphicx,color}
\usepackage{url}
\usepackage{amsmath}
\usepackage{booktabs}
\usepackage{multirow}
\usepackage{mathrsfs}
\usepackage{hyperref}

\usepackage{times}

\newenvironment{sciabstract}{%
\begin{quote} \bf}
{\end{quote}}

\newcounter{lastnote}

\title{Small vulnerable sets determine large network cascades in power grids}

\author{
Yang Yang,$^{1}$
Takashi Nishikawa,$^{1,2,\ast}$
Adilson E. Motter$^{1,2}$\\
\\
\normalsize{$^{1}$Department of Physics and Astronomy, Northwestern University, Evanston, IL 60208, USA}\\
\normalsize{$^{2}$Northwestern Institute on Complex Systems, Northwestern University, Evanston, IL 60208, USA}\\
\\
\normalsize{$^\ast$To whom correspondence should be addressed; E-mail: t-nishikawa@northwestern.edu.}
}

\date{}

\begin{document} 

\baselineskip18pt

%\noindent{\it \underline{Article Summary} --- see p.\ 4 for the full article}\\[4mm]
\noindent{\it \underline{Article Summary} --- followed by the full article on p.\ 4}\\[4mm]
\noindent{\Large\bf Small vulnerable sets determine large network cascades\\[2mm]in power grids}\\[6mm]
\noindent
Yang Yang,$^{1}$
Takashi Nishikawa,$^{1,2,\ast}$
Adilson E. Motter$^{1,2}$\\
\noindent Science {\bf 358}, eaan3184 (2017), \,\,
\href{https://doi.org/10.1126/science.aan3184}{DOI: 10.1126/science.aan3184}\\[2mm]
\noindent Animated summary: \url{http://youtu.be/c9n0vQuS2O4}\\[3mm]
{\small
$^{1}$Department of Physics and Astronomy, Northwestern University, Evanston, IL 60208, USA\\
$^{2}$Northwestern Institute on Complex Systems, Northwestern University, Evanston, IL 60208, USA\\
$^\ast$Corresponding author. E-mail: t-nishikawa@northwestern.edu
}

\vspace{7mm}

\noindent
Cascading failures in power
grids are inherently network processes, inwhich
an initially small perturbation leads to a sequence
of failures that spread through the
connections between system components. An
unresolved problem in preventing major blackouts
has been to distinguish disturbances that
cause large cascades from seemingly identical
ones that have only mild effects. Modeling and
analyzing such processes are challenging when
the system is large and its operating condition
varies widely across different years, seasons, and
power demand levels.

\bigskip\noindent
Multicondition analysis of cascade
vulnerability is needed to answer several
key questions: Under what conditions would
an initial disturbance remain localized rather
than cascade through the network? Which
network components are most vulnerable
to failures across various conditions? What is
the role of the network structure in determining
component vulnerability and cascade sizes?
To address these questions and differentiate
cascading-causing disturbances, we formulated
an electrical-circuit network representation
of the U.S.-South Canada power grid---a
large-scale network with more than 100,000
transmission lines---for a wide range of operating
conditions. We simulated cascades in
this system by means of a dynamical model
that accounts for transmission line failures
due to overloads and the resulting power
flow reconfigurations.

\baselineskip14pt

\noindent\includegraphics[width=\textwidth]{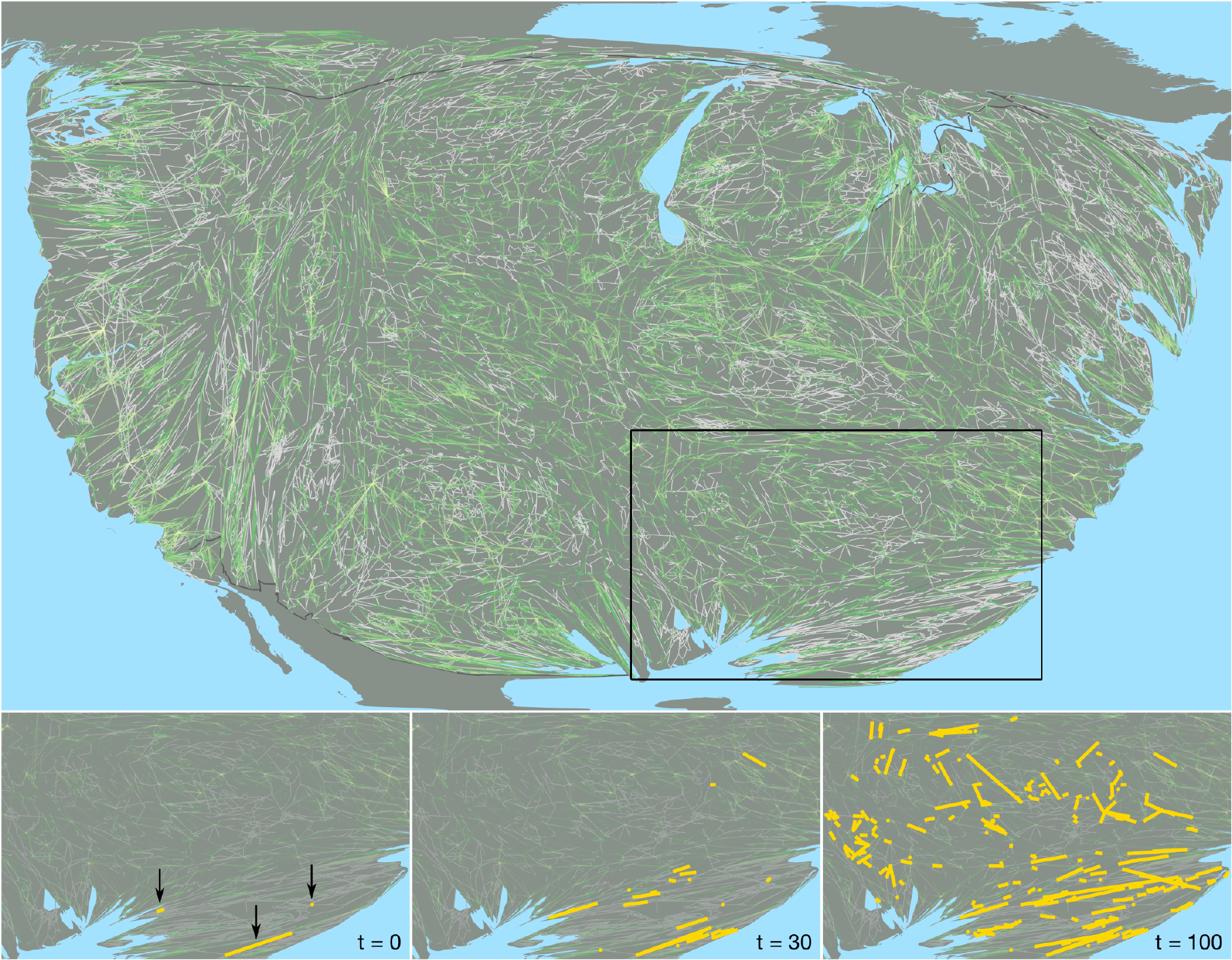}
\vspace{-3mm}

\noindent{Summary Figure: \bf Cascade-resistant portion of the U.S.-South Canada power grid.} The network is visualized on a cartogram that equalizes the density of nodes. (Top) Power lines that never underwent outage in our simulations under any grid condition are shown in green, whereas all the other lines---whose vulnerability varies widely---are in gray. (Bottom) Spreading of a cascade triggered by three failures at time $t = 0$ (arrows), which resulted in $254$ failures at $t = 100$ (the end of the cascade in linearly rescaled time).

\baselineskip18pt

\bigskip

\bigskip\noindent
To quantify cascade vulnerability,
we estimated the probability that each transmission
line fails in a cascade. Aggregating the
results from multiple conditions into a single
network representation, we created a systemwide
vulnerability map, which exhibits
relatively homogeneous geographical distribution
of power outages
but highly heterogeneous
distribution of the underlying
overload failures.
Topological analysis of
the network representation
revealed that the
transmission lines vulnerable to overload failures
tend to occupy the network's core, characterized
by links between highly connected
nodes. We found that only a small fraction of
the transmission lines in the network (well
below 1\% on average) are vulnerable under a
given condition. When measured in terms of
node-to-node distance and geographical distance,
individual cascades often propagate
far from the triggering failures, but the set
of lines vulnerable to these cascades tend to
be limited to the region in which the cascades
are triggered. Moreover, large cascades
are disproportionately more likely
to be triggered by initial failures
close to the vulnerable set.

\bigskip\noindent
Our results imply
that the same disturbance in a
given power grid can lead to disparate
outcomes under different
conditions---ranging from no damage
to a large-scale cascade. The association
between large cascades
and the triggering failures' proximity
to the vulnerable set indicates that
the topological and geographical
properties of the vulnerable set is
a major factor determining whether
the failures spread widely. Because
the vulnerable set is small, failures
would often repeat on the same lines
in the absence of interventions. Although
the power grid represents a
complex system in which changes
can have unanticipated effects, our
analysis suggests failure-based allocation
of resources as a strategy
in upgrading the system for improved
resilience against large cascades.

\maketitle 

\noindent{\bf One Sentence Summary:}
Computational analysis reveals that 
cascades in power grids
are driven by the recurrent failure of 
few---but central---components.

%:Abstract
\begin{sciabstract}
The understanding of cascading failures in complex systems has been hindered by the lack of realistic large-scale modeling and analysis that can account for variable system conditions. Here, using the North American power grid, we identify, quantify, and analyze the set of network components that are vulnerable to cascading failures under any out of multiple conditions. We show that the vulnerable set consists of a small but topologically central portion of the network and that large cascades are disproportionately more likely to be triggered by initial failures close to this set. These results elucidate aspects of the origins and causes of cascading failures relevant for grid design and operation, and demonstrate vulnerability analysis methods that are applicable to a wider class of cascade-prone networks.
\end{sciabstract}

%:Main text
\pagebreak
\noindent
Cascading failures are inherently large-scale network processes that cannot be satisfactorily understood from a local or small-scale 
perspective. In 
blackouts
caused by cascading failures in the power grid,
a relatively small local disturbance triggers a sequence of grid component failures, causing potentially large portions of the network to become 
inactive with costly outcomes.
In the North American power grid~\cite{grid_siz}, for instance, a single widespread power outage can inflict tens of billions of dollars in losses~\cite{LaCommare2006}, and smaller but more frequent outages can amount to a yearly combined impact comparable to that of the largest blackouts~\cite{hines2009large}. 
Yet,
not much is known about what distinguishes disturbances that cause cascades from seemingly identical ones that do 
not. Despite 
the significant advances made through conceptual modeling of general cascades~\cite{watts2002,goh_2003, crucitti_2004, motter_2004, kinney_2005, buldyrev_2010, brummitt_2012} and 
physics-based modeling of power-grid--specific cascades~\cite{anghel2007stochastic, manchesterModel2006, andrey:12, dobson2007complex},
a major 
obstacle still 
remains: the lack of realistic large-scale models and a framework for analyzing cascade vulnerability under variable system conditions. 
Developing such a framework is challenging for three reasons: (\textit{i}) detailed data combining both structural {\it and} dynamical parameters are scarce, (\textit{ii})
the system condition varies on a wide range of time scales, and (\textit{iii})~computational 
resources required for modeling grow combinatorially with system size~\cite{moore_2011}.
These challenges have
limited the applicability of most previous studies to vulnerability under a single condition and either to smaller scales than those required to describe large cascades or to models that are not constrained by real data.
Similar hurdles exist in studying large-scale failures in the broader context of complex networks~\cite{Strogatz2001,helbing2013,vespignani2009}, including extinction cascades in ecological systems~\cite{sahasrabudhe2011rescuing,eco:11,eco:12} and contagion dynamics in financial systems~\cite{haldane2011systemic,gai2010contagion}.

Here we focus on
the U.S.-South Canada power grid, which is the largest contiguous power grid amenable to  
modeling.
This system is composed of three interconnections (Texas, Western, and Eastern; Fig.~1A), which are separate networks of alternating current generators and power consumers connected by transmission lines (network components are illustrated in Fig.~1B).
To 
study
this system, we used the data reported in the Federal Energy Regulatory Commission (FERC) Form~715.
For each interconnection, 
the data represent
various \textit{snapshots} of the system, 
spanning the years 2008--2013 and
covering multiple 
seasons as well as both on- and off-peak demand levels, which correspond to different operating conditions.
Basic properties of the $46$ snapshots we used are listed in Table~S1.
A representation of each snapshot was constructed by processing
the parameters of individual power-grid components, including power generation and demand as well as the capacity of transmission lines.
Central to the analysis of cascade vulnerability in this system is that our approach distinguishes (\textit{i}) transmission lines (or simply lines) that do not carry flow because they have become out of service due to protective relay actions, equipment malfunctions, operational errors, or physical damages ({\it primary} failures); and (\textit{ii}) lines that do not carry flow at the end of the cascades because they are de-electrified due to the outage of other lines ({\it secondary} failures).

\section*{Geographic layout of vulnerabilities}

\noindent
The vulnerability of a given transmission line $\ell$ can be quantified by the probability $p_\ell$ that the line fails in a cascade event triggered by a random perturbation to a given snapshot of a given interconnection.
To estimate $p_\ell$, we used
a cascade dynamics model that combines key elements from previous models~\cite{OPA,paul:12,anghel2007stochastic} to suitably account for the physics of cascading failures.
In this model, the
initial state of the system for the given snapshot is determined by computing the power flow over all transmission lines and transformers from the 
power flow equation 
(see Materials and Methods, Supplementary Materials).
The triggering perturbation was implemented through
the removal of a set of $n_\text{t}$ lines, representing line outages due to unforeseen events, such as damage to power lines caused by extreme weather and unplanned line shutdowns caused by operational errors. 
After this initial removal, a cascade event was modeled as an iterative process, with each step consisting of a single power line outage due to 
overheating (primary failure), 
followed by the redistribution of power flow in the network to compensate for the lost flow over the failed 
line. 
Line overheating is modeled by a temperature evolution equation~\cite{anghel2007stochastic}, and flow redistribution is determined by solving 
the power flow equation again;
if
a primary line failure disconnects the network into multiple parts with 
unbalanced supply and demand, 
the power generation and consumption in each part are adjusted (similarly to how generation reserves and power shedding are used in grid operation) to allow for the subsequent power flow calculation.
The failure probability $p_\ell$ 
was
estimated from $K$ such simulated cascade events, including those with no subsequent failures.
Further details on the triggering perturbations and cascade dynamics model can be found in Materials and 
Methods, Supplementary Materials.

We validated the model against historical line outage data available from the Bonneville Power Administration (BPA) with respect to the distribution of cascade sizes measured by the number of (primary) line failures $N_\text{f}$ (Materials and 
Methods, Supplementary Materials, and
Fig.~S1A).
We also validated the extremal cascade size measured by $N_\text{f}$ and power shed $P_\text{s}$ (defined as the reduction in the amount of power delivered to the consumers) against the BPA data and grid disturbance
data from the North American Electric Reliability Corporation (NERC), respectively (Materials and 
Methods, Supplementary Materials, and Fig.~S1, B and C).
All simulations were performed with $n_\text{t}=3$, as the cascade size distribution for a given snapshot did not differ significantly for other choices of $n_\text{t}$ (Fig.~S2).
However, the distribution exhibited considerable variation across different snapshots,
both when cascade size was 
measured by the power shed $P_\text{s}$ (Fig.~S3), and when measured by the number of line failures $N_\text{f}$ (Fig.~S4).

To aggregate results over different snapshots, we used
a {\it node} to represent 
the set of all buses associated with the same geographic location across all snapshots in our dataset, where the term bus refers to 
a connection point between components of a power grid, such as transmission lines, transformers, and generators (Fig.~1B).
This definition of a node typically corresponds to a substation 
and can include generators at a nearby power plant and/or an electrical load representing local power consumption. 
We used a \textit{link} between a pair of
nodes to represent
the set of all 
(parallel)
transmission lines directly connecting the same pair of nodes
in at least one 
snapshot, 
where each of these transmission lines
connects two different 
buses (one from each node in the pair).
In this network, the aggregated vulnerability $p_l := \langle p_{\ell} \rangle $ of a link $l$, which we refer to as the {\sc a}-\textit{vulnerability}, is a weighted average of the failure probabilities over the lines represented by the link $l$ and over the various snapshots:
\begin{equation}
 \langle p_{\ell} \rangle  =\frac{\sum_{c}\sum_{\ell} p_{\ell; c} w_{c}} {\sum_{c}\sum_{\ell} w_{c}},
 \label{vuln}
\end{equation}
where $c$ indexes the different snapshot conditions simulated, and the sum over $\ell$ is limited to the set of transmission lines defining the link $l$ for the given $c$. 
Here, $p_{\ell; c}$ is the probability of line failure in the simulated perturbations of the system (the values of $K$ we used are given in Table~S1 and justified in Fig.~S5)
and $w_c$ represents the weight assigned to each snapshot 
(Table~S1).
In our analyses, we present the {\sc a}-vulnerability separately  for primary failures $\bigl($denoted by $\langle p^{\text{(p)}}_{\ell} \rangle \bigr)$, secondary failures $\bigl($denoted by $\langle p^{\text{(s)}}_{\ell} \rangle \bigr)$, and the combination of both primary and secondary failures (denoted by $\langle p_{\ell} \rangle $ itself). 

We constructed the {\sc a}-vulnerability map of the U.S.-South Canada power grid (shown in 
Fig.~2, A to C, 
for a
portion of the grid).
Over the entire network, we found that 
only $10.8\%$ of all links ever underwent
a primary failure in our simulations and that secondary failures 
were on average $3.77$ times more prevalent than primary 
ones (Table~S3).
We also found that 
{\sc a}-vulnerability was
very unevenly distributed 
among the links,
with 20\% of the failing links (which in the case of primary failures correspond to only 2.16\% of all links) accounting for 
about 85\%, 66\%, and 69\% of 
the primary failures,  secondary failures, and  combined set of all failures, 
respectively (Fig.~2, D to F).
Also uneven was the geographical distribution of links with nonzero {\sc a}-vulnerability (Fig.~2, A to C), whose density was correlated positively with population density.
This correlation was
mainly due to the bias toward higher density of links in more densely populated areas, as
it disappeared
when {\sc a}-vulnerability was
averaged over the links in each geographical area to 
control for this bias.
However, substantial
geographical heterogeneity still remained
for the averaged {\sc a}-vulnerability, ranging over several orders of magnitude when calculated for individual U.S.\ counties.
These observations were validated with
the U.S.\ county population data from the 2010 census and the geographic coordinates of county boundaries (Fig.~S6).
Among the $48$ states and the District of Columbia represented
in the U.S.\ portion of the network, the three least vulnerable ones 
were West Virginia (average $\langle p_\ell\rangle$ of $3.2 \times 10^{-5}$), Tennessee (average $\langle p_\ell\rangle$ of $3.5 \times 10^{-5}$), and Mississippi (average $\langle p_\ell\rangle$ of $3.8 \times 10^{-5}$), all in the middle third of the population density ranking.
However, some states among the least vulnerable did
have relatively high or low population density, such as Illinois and Nebraska, which ranked
$13$th and $43$rd in population density while having the $5$th and $6$th lowest {\sc a}-vulnerability, respectively. 
The heterogeneity of {\sc a}-vulnerability is visualized in Fig.~3A 
with
a map representation that equalizes the density of nodes.
The breakdown of this representation into primary and secondary failures, presented in 
Fig.~3, B and C, 
shows that {\sc a}-vulnerability to primary failures 
was more heterogeneously distributed than {\sc a}-vulnerability to secondary failures.  
Over all pixels with nonzero {\sc a}-vulnerability, the standard deviation of $ \log\,  \langle \overline{ p^{\phantom{\text{p}}}_\ell }\rangle $ was $0.48$ ($89.2\%$), of $ \log\, \langle \overline{p^{\text{(p)}}_\ell} \rangle  $ was $0.58$ ($57.5\%$) and of $ \log \, \langle \overline{p^{\text{(s)}}_\ell} \rangle $ was $0.41$ ($87.0\%$), where the number in parenthesis represents the fraction of such pixels.  
The homogeneity in the distribution of secondary failures, which were several times more numerous than primary failures, underlies the relatively homogeneous aggregated distribution of the resulting power outages observed in Fig.~3A.

\section*{Network characterization of vulnerabilites}
Our characterization of {\sc a}-vulnerability allows us to 
study how 
the observed cascade dynamics depend 
on the network 
structure and to identify the topological centrality of individual links as a determinant. 
Topological 
centrality can be quantified through the 
concept
of $k$-core~\cite{kcore-orig-1,kcore-orig-2,kcore,vis_kcore}, which is defined as the largest subnetwork in which every node has at least $k$ links (that is, it has degree $k$).
The $k$-core of a given network can be obtained by recursively removing all nodes with degree $<k$ until all nodes in the remaining network have degree $\ge k$.
Repeating this for $k=1,2,\ldots$ determines the $k$-core decomposition of the network.
The coreness of a node is then defined as the (unique) integer $c$ for which this node belongs to the $c$-core but not to the $(c+1)$-core~\cite{graph_theory}.
We further extend this concept to links by defining a link's coreness to be the smaller coreness of the two nodes it connects.
Figure~4A illustrates a network visualization based on this decomposition.

When this network decomposition 
was
applied to the entire topology of the U.S.-South Canada power system, we 
found
that links of coreness $2$ were
dominant in all three interconnections (with $81\%$, $67\%$, and $82\%$ 
of all links in the Texas, Western, and Eastern networks, respectively).
This dominance of coreness $2$ links was also observed for
the cascade-prone portion of the network and was further verified
separately for the set of links vulnerable to primary failures $\bigl(\langle \overline{p^{\text{(p)}}_\ell} \rangle > 0 \bigr)$ as well as the set of links vulnerable to secondary failures $\bigl(\langle \overline{p^{\text{(s)}}_\ell} \rangle > 0 \bigr)$.
These results are visualized in Fig.~4B
for the case of the Eastern interconnection.

Upon closer inspection, however, the vulnerability revealed
a strong correlation with link coreness beyond what can be inferred from the availability of links of a given coreness in the network.
For primary failures, 
almost all links of coreness $1$ showed
zero {\sc a}-vulnerability in our simulations, 
whereas $7$ to $19$\%
of higher coreness links were
vulnerable (Fig.~4C).
The links of coreness $1$ are rarely
vulnerable because each belongs to a tree subnetwork connected to the rest of the network through a single node, and this protects the link from flow rerouting, which is responsible for most primary failures (e.g., flow rerouting 
accounts for more than $98$\% of primary failures 
in
the $2010$ spring peak snapshot of the Texas network, as shown in 
Supplementary Materials, Materials and Methods, ``Identifying mechanisms responsible for primary failures'').
Among the links that were vulnerable, the level of {\sc a}-vulnerability increased 
monotonically with their coreness (Fig.~4D).
This is probably because there are more flow paths (from power generators to consumers) that are parallel to a link of higher coreness in general, making the link more likely to be affected by flow rerouted from a failure in these paths.

For secondary failures, the fraction of links that were
vulnerable and the {\sc a}-vulnerability levels of these links followed
opposite trends.
The decrease in the fraction of vulnerable links shown in 
Fig.~4E can be understood by noting that a link can experience
a secondary failure only if all available flow paths passing through that link are disabled by primary failures.
As links of higher coreness generally have more such paths, they were
less likely to fail through this mechanism.
Among the vulnerable links, the increase of the average {\sc a}-vulnerability
with coreness shown in Fig.~4F likely arose from 
the organization of the nodes in each $k$-core into graph components
(maximal subsets of nodes in which every node pair is connected by a network path). 
Whereas the 
$2$-core formed
a single 
graph component
in all three interconnections, 
the nodes in
the $3$-core were
organized into multiple
graph components 
($3$, $11$, and $52$ components for the Texas, the Western, and the Eastern network, 
respectively), which were
connected sparsely with each other by coreness $2$ links.
Because of this structure, most secondary failures on links of coreness $\ge 3$ 
were likely caused by primary failures on the surrounding links of coreness $2$ that disconnected
a $3$-core graph component with no internal power generation from the 
other $3$-core components.
This would make 
the links in 
these components
prone 
to repetitively
undergo secondary failures together.
This tendency of co-occuring failures~\cite{Yang2017} among vulnerable links 
would lead to higher {\sc a}-vulnerability for those links than for links with lower coreness.

\section*{Relating triggers and network states to vulnerable lines}

\noindent
To characterize the lines at risk of primary failures,
we
now shift our attention back to individual transmission lines connecting buses in each snapshot, rather than their collective representation as 
links. 
For this purpose,
we define a \textit{vulnerable} transmission line for a given snapshot t			o be a line $\ell$ for which $p^{\text{(p)}}_\ell > 0.0005$ with at least 95\% Wilson's confidence level~\cite{brown2001interval} (which excludes any line with a single failure in $1{,}000$ simulated events).
This approach for vulnerability analysis is in contrast to previous studies on identifying the line failure combinations that initiate large cascading failures (i.e., a single snapshot)~\cite{paul:12, long:16}.
We then define the \textit{vulnerable set} $\mathcal{V}$ to be the set of all vulnerable lines for the given snapshot.
We 
found
that these vulnerable sets not only represented
small portions of the grid in each snapshot but also exhibited considerable
overlap across different snapshots (although it was rare  
for the same line to be
vulnerable in all snapshots).
These findings are 
presented
in 
Table~1 
for each interconnection using, respectively, 
the weighted average $\langle|{\cal V}|\rangle$ of the number of vulnerable transmission lines over all snapshots
and the number $|{\cal V}_{\cap}|$ of lines that were
vulnerable in two or more 
snapshots (relative 
to the number expected if the vulnerable sets were 
randomly distributed with no correlation).
For example, in the Texas interconnection, $\langle|{\cal V}|\rangle=48$ represents only about $0.6\%$ of all the transmission lines and the relative number of overlapping lines $|{\cal V}_{\cap}|$ is $2.9$.
(Details on 
the distribution of $p^{\text{(p)}}_\ell$
for individual snapshots can be found in Fig.~S7.)

Having a small portion of the grid vulnerable to cascading failures does not imply that these failures stayed localized even for single snapshots.
To quantify the degree to which cascades were
localized, we used
the 
concepts
of 
topological distance (the number of links along the shortest paths in the network) and geographical distance (the arc length along the Earth's surface), both normalized by the size of the triggering region measured by the respective distances and thus are unitless (Materials and 
Methods, Supplementary Materials).
Specifically, 
the extent of the vulnerable set was measured
by 
$d_{\text{v-v}}$ and $g_{\text{v-v}}$, defined as the normalized topological and geographical
distance, respectively,
between two transmission lines, 
averaged over all pairs of lines in the vulnerable set.
We further defined $\langle d_\text{v-v} \rangle$ and $\langle g_\text{v-v} \rangle$ to be the weighted average of $d_\text{v-v}$ and $g_\text{v-v}$, respectively, over all snapshots.
Table~1 
shows that, for the Texas and 
Western networks, 
both $\langle d_\text{v-v} \rangle$ and $\langle g_\text{v-v} \rangle$ are
comparable with the size of the interconnection, revealing 
that the spreading of cascades is nonlocal
[which is consistent with observations 
from historical data~\cite{dobson_nonlocal_2016}, from power flow calculations~\cite{Jung_long_2015}, and from abstract models~\cite{daqing:14,nonlocal_spread}].
In
all cases 
$\langle d_\text{v-v} \rangle<1$ and $\langle g_\text{v-v}\rangle<1$ 
hold true in the Eastern interconnection, where cascades were
actually triggered in a local region and could have,
in principle, spread 
widely to the other regions within the 
interconnection, leading to
$\langle d_\text{v-v} \rangle>1$ or $\langle g_\text{v-v} \rangle >1$. 
This suggests that there is also an aspect of the cascading failures that is local: the propagation of failures in general does not extend too far from the region being perturbed.

The analysis of vulnerable sets provide relevant insights not only into the origins of cascading failures, but also into the size of the damage inflicted on the network by individual cascades.  
In particular, what is the difference between the perturbations that cause large cascades and those that do not?
To answer this question quantitatively, 
we categorized
cascades according to their sizes measured by the power shed $P_\text{s}$ defined above: 
small cascades ($0.01\text{MW} \leq P_\text{s}<300$MW) and large cascades ($P_\text{s} \geq 300$MW).
This choice of measure and threshold is based on the 
NERC
requirement that all blackouts causing more than $300$MW of lost power be reported.
We characterized
perturbations by three different measures based on (normalized) distances: $d_{\text{t-t}}$, defined as the average pairwise distance among the $n_\text{t}$ triggering line failures, as well as $d_{\text{t-v}}$ and $g_{\text{t-v}}$, defined as the minimum topological and geographical distances, respectively, from one triggering line failure to the vulnerable set $\mathcal{V}$.
Figure~5 
shows the average of these distances ($\bar{d}_{\text{t-t}}$, $\bar{d}_{\text{t-v}}$, and $\bar{g}_{\text{t-v}}$) over cascades in each size category for each region.
Cascades 
resulting in power shed $P_\text{s} \geq 300\text{MW}$ were
associated with a set of triggering line failures that were
topologically closer to each other (Fig.~5A), 
as well as with triggering failures that occurred
topologically and geographically closer to a vulnerable line (Fig.~5, B and C).

\section*{Conclusions}

\noindent
Our vulnerability analysis of a continent-wide power system distinguishes itself from most previous studies by its scale, but also by accounting for: (\textit{i}) the physics of cascading failures (DC-approximated power flow redistribution and heating of line conductors); (\textit{ii}) grid operation practices (generation reserves and power shedding); and (\textit{iii}) a wide range of conditions across years, seasons, and power demand levels (over which the average cascade size varies by one to two orders of magnitude). A strength of our approach is that it consists of tools---the definition of vulnerable sets, the method for aggregating multiple network conditions, and the analysis of coreness-vulnerability correlations---that are applicable to any cascade-prone network.

Our analysis separates the set of all failures occurring in cascade events into  primary failures, which define the vulnerable set and account for only 1/5 of all failures, and secondary failures, which are more uniformly distributed and, albeit more numerous, are a mere consequence of the primary ones. 
The vulnerable set is not only surprisingly small but also highly skewed---with few lines far more likely to undergo a primary failure than the others---and patchy even when we control for the heterogeneity in the geographic organization of the grid.
{Although the vulnerable set is widespread through the network, the portion of it recruited in each cascade is not, and is in fact} strongly spatially correlated with the location of the triggering line failures; this is counter to the perception that cascades [for being nonlocal with respect to both topological and geographical distances~\cite{nonlocal_spread3,Yang2017}] can spread essentially without spatial constraints.

Our analysis also shows that larger cascades are associated with co-occurring perturbations that are closer both to each other and to the vulnerable set.
This validates the existing hypothesis that localized triggering failures amount to bigger cascades~\cite{local_triggers} and reveals a striking relation to the classic threshold model~\cite{watts2002} used to describe behavioral cascades in social systems, where large cascades tend to be triggered by perturbations adjacent to the set of ``early adopters.'' 
This set corresponds to the nodes most susceptible to change and thus plays a role similar to the one the vulnerable set plays in our analysis.
The network topology emerged as a significant factor in determining the risk of cascading failures in our analysis 
based on
the $k$-core decomposition, which has also been used to characterize 
nodes that serve as efficient spreaders
in 
contact-based 
processes~\cite{Kitsak:2010}.

There are never two identical cascades in a network.
It may thus come as a surprise that (primary) failures in large cascades are constrained to only a small subset of the network, which will likely experience new failures in the absence of remediating actions. 
This offers a scientific foundation for {\it failure-based allocation of resources}, which in the case of a power grid would be based on prioritizing upgrades of the system on the basis of previous observed failures~\cite{dobson2007complex}---but only if those are the primary (as opposed to {\it all}) failures (although upgrading transmission line capacities in the vulnerable set could create new vulnerable lines outside the set).
Future work will be needed to determine the extent to which this applies to
other flow networks that are subject to repeated failures, such as supply chains, food webs, and traffic networks.

\section*{Methods summary}

\noindent
For each interconnection, the system was modeled as a network of buses connected by transmission lines, given the parameters of individual network components in a given snapshot.
The triggering perturbations were chosen uniformly from all lines for the Texas and Western networks, whereas for the Eastern network, they were chosen uniformly within one of the six regions defined by NERC (Fig.~1A and Table S2).
The initial state of the network and the redistribution of power flow following a line removal were both calculated by solving an equation that expresses a balance between incoming and outgoing power flows at each bus.
Through a temperature evolution equation, the heating of a transmission line was modeled as an exponential convergence to the equilibrium temperature determined by the power flow over that line.
Mechanisms responsible for the primary failures occurring in a given simulated cascade were identified using an algorithm we developed to determine the degree to which the change in each generator's output contribute to changes in individual line power flows.

The density-equalizing transformation used to generate Fig.~3 was determined by estimating the density function for the geographical distribution of nodes and evolving it to a uniform-density equilibrium through a linear diffusion process~\cite{gastner2004diffusion}.
The topological and geographical distances between two transmission lines are defined based on the corresponding distances between the buses they connect.
Both distances are thus zero between two lines that connect to a common bus.
Further details on the formulation of the power flow equation, triggering perturbations, temperature evolution equation, validation of the cascade dynamics model against historical data, calculation of the density-equalizing transformation, algorithm for assigning power flow changes to generators, and the definitions of bus-to-bus distances, are all given in the supplementary materials.

\pagebreak

%:References
\bibliographystyle{Science}

\pagebreak

%:Acknowledgments
\section*{Acknowledgments}

\noindent
The authors thank Hamed Valizadehhaghi for insightful 
discussions.
This work was supported by 
the Institute for Sustainability and Energy at Northwestern (ISEN) under a Booster Award, 
the U.S. National Science Foundation under Grant 
DMS-1057128, and
the Advanced Research Projects Agency--Energy (U.S. Department of Energy) under Award Number DE-AR0000702. 
The views and opinions of authors expressed herein do not necessarily state or reflect those of the United States Government or any agency thereof.
The power-grid data
were 
obtained
from FERC 
under
a non-disclosure agreement by following the procedure described at \url{https://www.ferc.gov/legal/ceii-foia/ceii.asp}.
The BPA line outage data and the NERC 
grid disturbance
data are both publicly available at \url{https://transmission.bpa.gov/Business/Operations/Outages/} (Miscellaneous Outage Data and Analysis) and \url{https://www.oe.netl.doe.gov/OE417_annual_summary.aspx} (Electric Disturbance Events, OE-417), respectively.
The 2010 U.S.\ census data and the boundary data for the U.S.\ counties can be downloaded from \url{https://factfinder.census.gov/} and 
\url{https://www.census.gov/geo/maps-data/data/cbf/cbf_counties.html}, 
respectively.

\section*{Supplementary materials}
Materials and Methods\\
Figs. S1 to S7\\
Tables S1 to S3

%:Tables
\clearpage

\noindent\textbf{Table 1. Subdivisions of the U.S.-South Canada power grid 
and its vulnerable sets.}
The rows represent the regions defined by 
NERC (Fig.~1A and Table~S2),
within which the simulated cascades are 
triggered.
The columns represent the number of buses, number of transmission lines, and four measures of the vulnerable sets: 
the number of vulnerable lines $|{\cal V}|$, the relative number of lines that are vulnerable in multiple
snapshots $|{\cal V}_{\cap}|$, and the mean pairwise 
normalized
topological and geographical 
distances between vulnerable lines, $d_\text{v-v}$ and $g_\text{v-v}$, respectively.
These quantities are averaged over all snapshots $\bigl($which is indicated by the notation $\langle\, \cdot\, \rangle\bigr)$.  The  
normalized
distances are defined in Materials and Methods.
\bigskip
\begin{center}
\begin{tabular}{lrrp{1pt}rrrr}
\hline\\[-11pt]
 \multicolumn{3}{c}{Interconnections} & &
 \multicolumn{4}{c}{Vulnerable sets} \\[2pt]
\cline{1-3}
\cline{5-8}\\[-11pt]
& $\langle$Buses$\rangle$ & $\langle$Lines$\rangle$ & & $\langle|{\cal V}|\rangle$ & $|{\cal V}_{\cap}|$ & $\langle d_\text{v-v} \rangle$ &$\langle g_\text{v-v} \rangle$ \\[3pt]
\hline\\[-11pt]
Texas & $6{,}161$ & $7{,}637$  & & $48$ &  
$2.9$ & $0.82$ & $0.70$ \\
Western & $15{,}891$ & $20{,}397$   & & $81$ &  
$5.9$ & $0.84$ & $0.95$\\
Eastern & $56{,}740$ & $72{,}903$\\
\quad FRCC & &   
                     & &   $37$
                     & $1.1$ & $0.69$ & $0.70$ \\
\quad MRO & & 	
                   & &   $32$
                   & $3.4$ & $0.79$ & $0.97$ \\
\quad NPCC & & 
                   & &   $130$
                   & $2.1$ & $0.85$ & $0.72$\\
\quad RFC & & 
                  & &   $76$
                  & $4.5$ & $0.94$ & $0.91$\\
\quad SERC & & 
                  & &   $11$
                  & $11.6$ & $0.92$ & $0.94$ \\
\quad SPP & & 
                  & &  $14$
                    & $3.3$ & $0.66$ & $0.63$\\[2pt]
%& $\langle$Buses$\rangle$ & $\langle$Lines$\rangle$ & & $|{\cal V}_{\cap}|$ &$\langle|{\cal V}|\rangle$& $\langle \bar{d}_\text{v-v} \rangle$ &$\langle \bar{g}_\text{v-v} \rangle$ \\[3pt]
%\hline\\[-11pt]
%Texas & $6{,}161$ & $7{,}637$  & & $2.9$ &  
%$48$ & $0.82$ & $0.70$ \\
%Western & $15{,}891$ & $20{,}397$   & & $5.9$ &  
%$81$ & $0.84$ & $0.95$\\
%Eastern & $56{,}740$ & $72{,}903$\\
%\quad FRCC & &   
%                     & & $1.1$  
%                     & $37$ & $0.69$ & $0.70$ \\
%\quad MRO & & 	
%                   & & $3.4$  
%                   & $32$ & $0.79$ & $0.97$ \\
%\quad NPCC & & 
%                   & & $2.1$  
%                   & $130$ & $0.85$ & $0.72$\\
%\quad RFC & & 
%                  & & $4.5$  
%                  & $76$ & $0.94$ & $0.91$\\
%\quad SERC & & 
%                  & & $11.6$  
%                  & $11$ & $0.92$ & $0.94$ \\
%\quad SPP & & 
%                  & & $3.3$ 
%                    & $14$ & $0.66$ & $0.63$\\[2pt]
\hline
\end{tabular}
\end{center}

%:Figures
\clearpage

%:Fig 1
\begin{figure*}[t!]  
\vspace{-15mm}
\begin{center}
\includegraphics[width=1.0\textwidth]{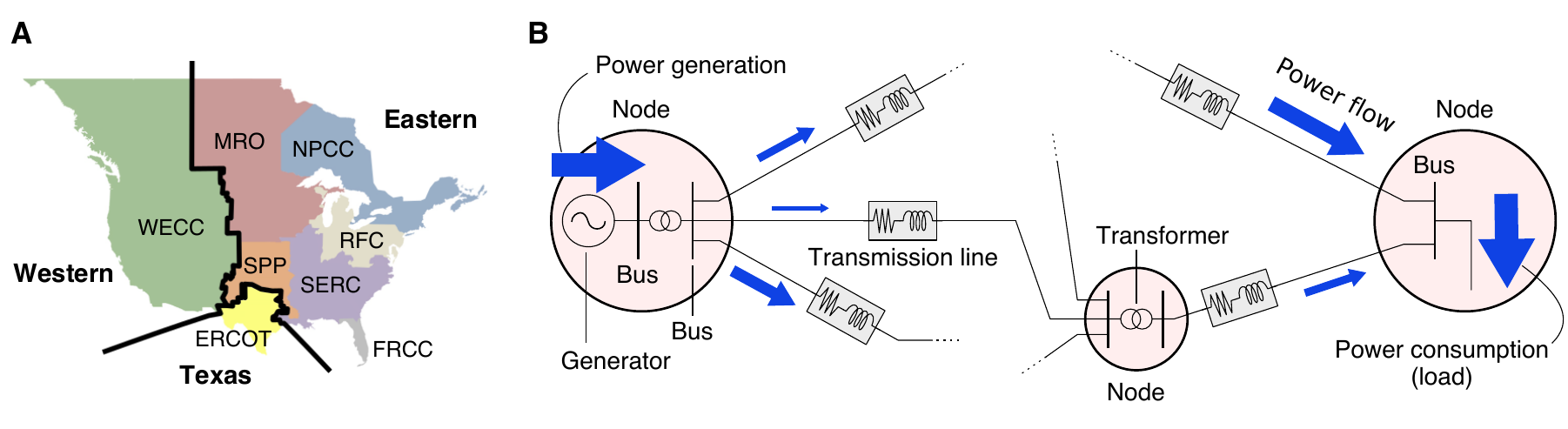}
\caption{
\textbf{The U.S.-South Canada power transmission network.}
(\textbf{A})~System map showing the Texas, Western, and Eastern interconnections, as well as the eight NERC regions (Table~S2).
(\textbf{B})~Schematic diagram of a portion of a transmission network.
The vertical lines and pink circles represent buses and nodes, respectively.
As indicated by the blue arrows, power injected by the generators flows through this network of transmission lines and is eventually consumed at other points (where the thickness of the arrow represents the amount of power flow).}
\end{center}
\end{figure*}

%:Fig 2
\begin{figure*}[t!]  
\vspace{-15mm}
\begin{center}
\includegraphics[width=1.0\textwidth]{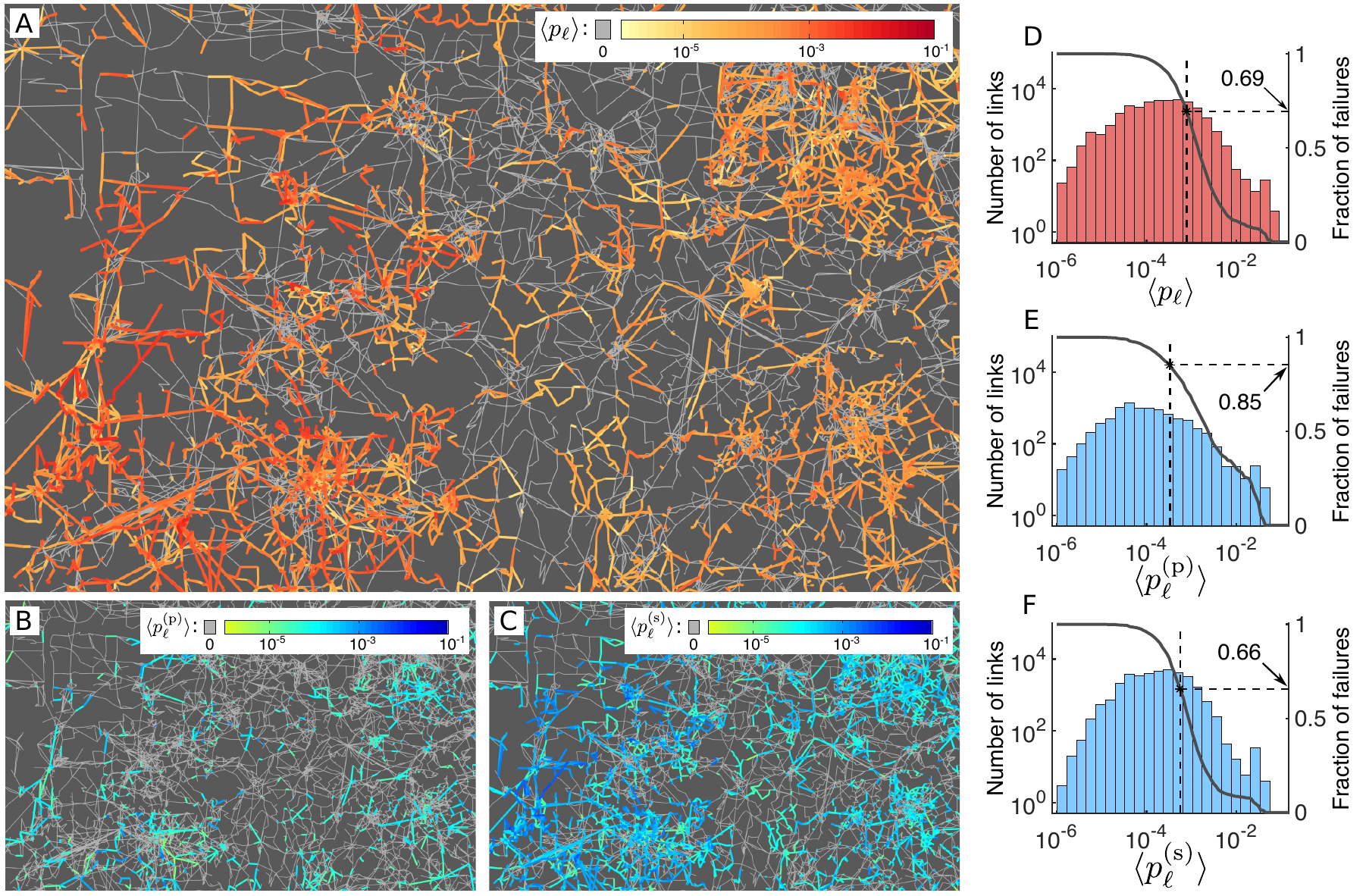}
\caption{
\textbf{Vulnerability map of the U.S.-South Canada power grid.}
(\textbf{A}) 
Averaged failure probability of transmission lines, including primary and secondary failures, expressed as the total
{\sc a}-vulnerability of links $\langle p_\ell \rangle $.
Since the structure of the power grid varies slightly from one snapshot to another, we visualize the {\sc a}-vulnerability using 
a single network constructed to represent all snapshots of each interconnection by regarding 
the set of all buses at a given geographical location
as a node and all transmission lines connecting two 
nodes
as a single link. 
Each link is color-coded by the failure probability $\langle p_\ell \rangle$ estimated as a weighted average over all lines in all 
snapshots, where gray indicates links 
whose estimated probability is zero.
(\textbf{B} and \textbf{C})
Same as in (A), but color-coded separately for the {\sc a}-vulnerability to primary failures $\langle p^{\text{(p)}}_\ell \rangle $~(B) and the {\sc a}-vulnerability to secondary failures $\langle p^{\text{(s)}}_\ell \rangle $~(C).
Panels (A), (B), and (C) correspond to the same unidentified portion of the U.S.-South Canada map.
(\textbf{D}) Histogram of the {\sc a}-vulnerability
$p_l = \langle p_\ell \rangle$ 
and the curve for
$f(p) := \sum' p_l / \sum_{l} p_l$, where $\sum'$ denotes the sum over all links $l$ satisfying $p_l \ge p$.
The function $f(p)$ thus represents the fraction of all failures that are associated with links of {\sc a}-vulnerability $p$ or larger.
(\textbf{E} and \textbf{F}) Same as in (D), but with $\langle p_\ell \rangle $ replaced by $\langle p^{\text{(p)}}_\ell \rangle $ and $\langle p^{\text{(s)}}_\ell \rangle $, respectively.
In each of the panels (D), (E), and (F), the
vertical and horizontal dashed lines indicate, respectively, the minimum {\sc a}-vulnerability $p^*$ among the most vulnerable $20$\% of all failing links and the fraction $f(p^*)$ of failures accounted for by these links.}
\end{center}
\end{figure*}

%:Fig 3
\begin{figure*}[t!]  
\begin{center}
\includegraphics[width=4.1in]{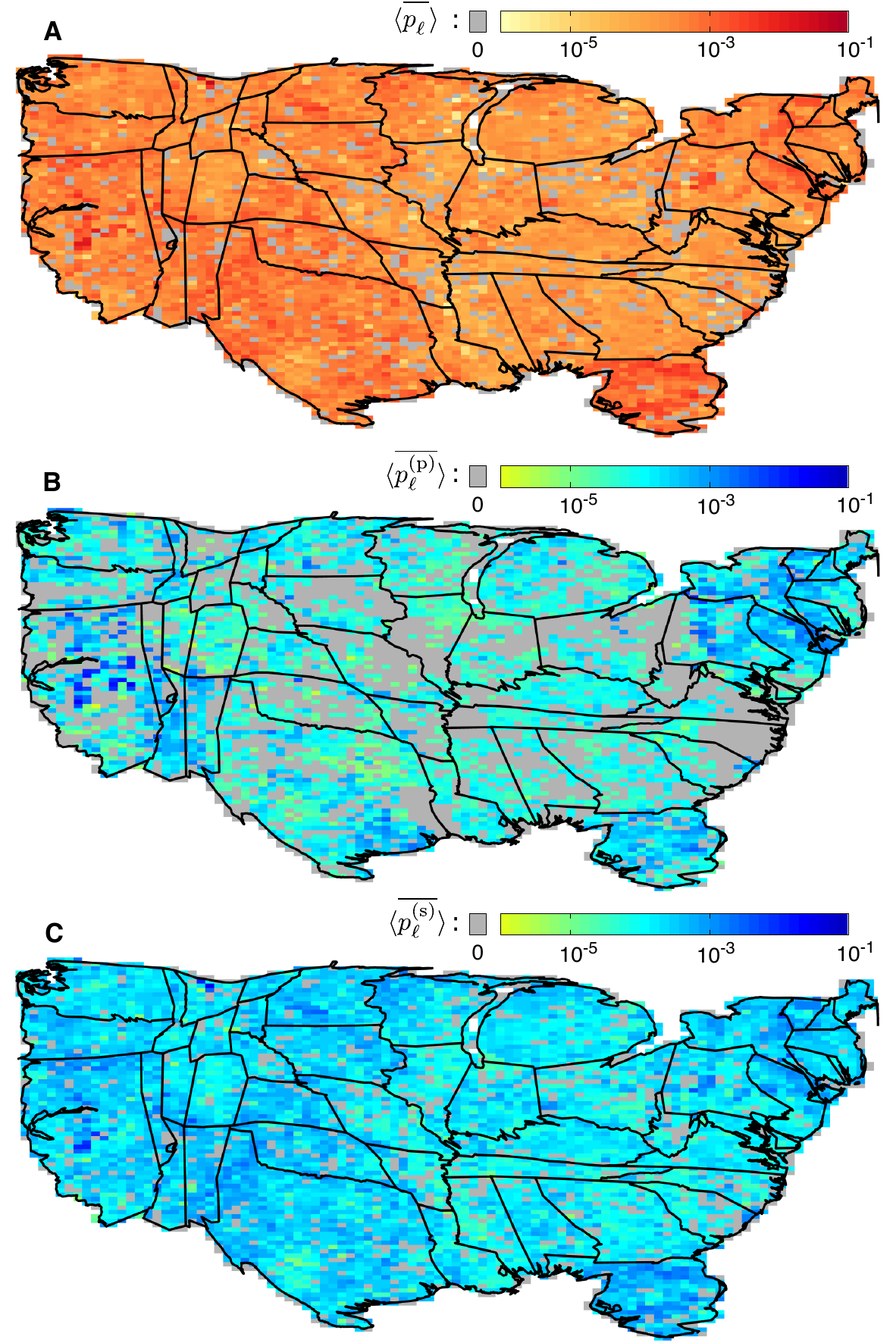}
\caption{
\textbf{Vulnerability of the power grid on a density-equalizing map.} 
(\textbf{A}) Each pixel is color-coded by the average {\sc a}-vulnerability 
$\langle \overline{p^{\protect\phantom{\text{p}}}_\ell}\rangle$, 
including both primary and secondary failures,  over all links connected to 
nodes
in the area of  the pixel.  
The cartogram 
was
generated using the diffusion-based method in Ref.~\cite{gastner2004diffusion} to equalize the 
density of nodes
(Materials and Methods), and is limited to the U.S. portion of the 
network.
Color gray marks the pixels with zero average {\sc a}-vulnerability.
(\textbf{B} and \textbf{C}) Same as in (A) but color-coded separately for the average {\sc a}-vulnerability to primary failures  $\langle \overline{p^{\text{(p)}}_\ell} \rangle $ (B) and the average {\sc a}-vulnerability to secondary failures $\langle \overline{p^{\text{(s)}}_\ell} \rangle $ (C).
}
\end{center}
\end{figure*}

%:Fig 4
\begin{figure*}[t!]  
\begin{center}
\includegraphics[width=5.1in]{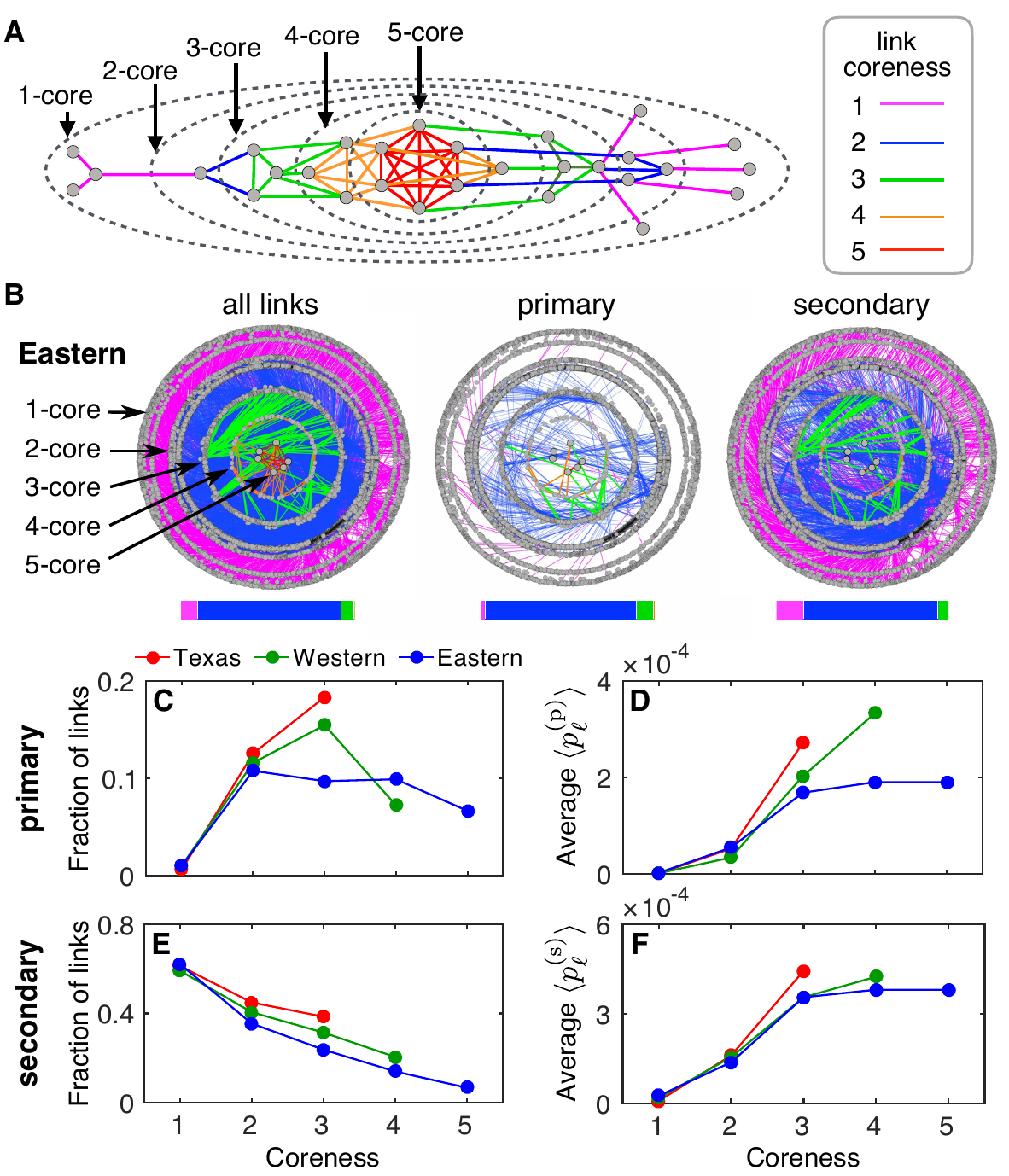}
\caption{
\textbf{Characterization of vulnerability through $k$-core decomposition.} 
(\textbf{A}) Coreness-based network visualization, where nodes with higher coreness are placed closer to the center.
(\textbf{B}) Visualization of the $k$-core decomposition of the Eastern interconnection, showing (left) all the links in the network, (middle) only the links with nonzero {\sc a}-vulnerability to primary failures $\bigl(\langle p^{\text{(p)}}_\ell \rangle>0\bigr)$, and (right) only the links with nonzero {\sc a}-vulnerability to secondary failures $\bigl(\langle p^{\text{(s)}}_\ell \rangle>0\bigr)$.
The bar under each panel shows the distribution of link coreness,
color-coded as in (A).
(\textbf{C}) Fraction of links with $\langle p^{\text{(p)}}_\ell \rangle>0$ among all links of a given coreness.
(\textbf{D})~Average of $\langle p^{\text{(p)}}_\ell \rangle$ over all links of a given coreness with $\langle p^{\text{(p)}}_\ell \rangle>0$.
(\textbf{E} and \textbf{F}) Counterparts of (C) and (D) for secondary failures.}
\end{center}
\end{figure*}

%:Fig 5
\begin{figure*}[t!]  
\begin{center}
\includegraphics[width=5in]{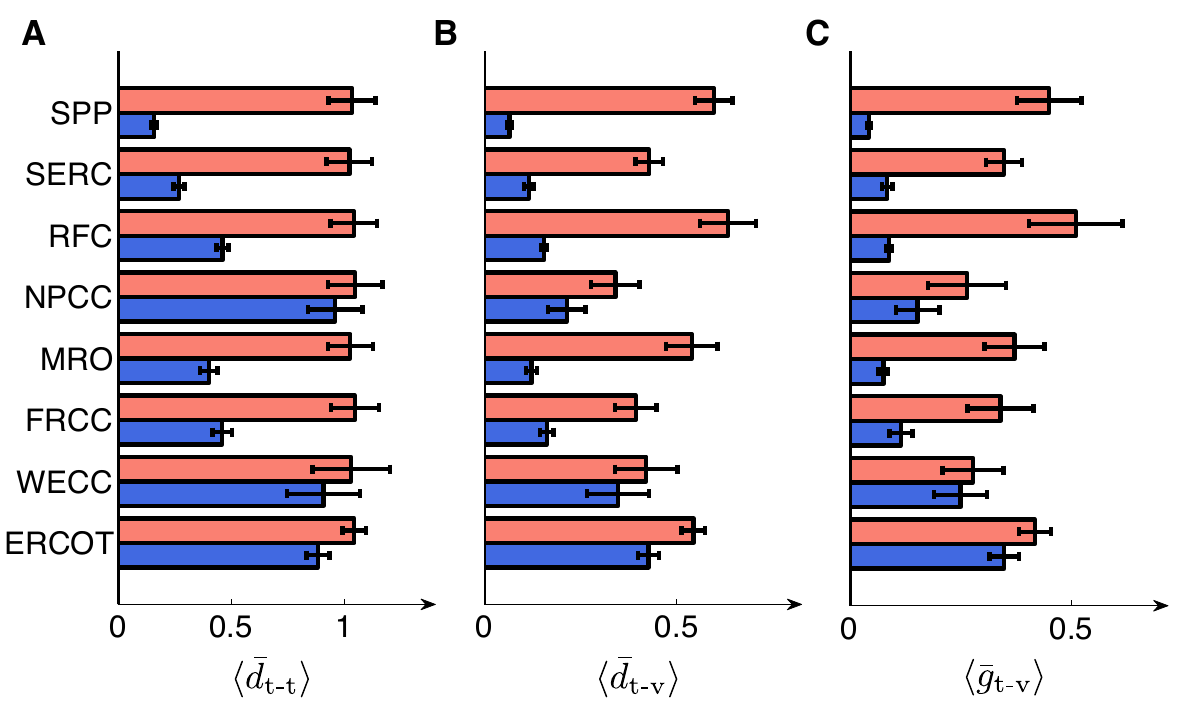}
\caption{
\textbf{Cascade size and distances involving triggering line failures.}
(\textbf{A} to \textbf{C}) Three types of (normalized)
distances are shown for each NERC region (Fig.~1A and Table S2): 
the mean pairwise topological distance between the triggering failures (A),
the topological distance between the set of triggering failures and the vulnerable set (B), and
the geographical distance between the set of triggering failures and the vulnerable set (C).
The distances are averaged separately over large cascades (blue, $P_\text{s} \geq 300\text{MW}$) and over small cascades (red, $0.01\text{MW} \leq P_\text{s}<300\text{MW}$).
In each case, the distances are further averaged over all snapshots.
Error bars mark the estimated standard deviation.}
\end{center}
\end{figure*}

%:Supplementary Materials
\clearpage
\baselineskip18pt

\setcounter{equation}{0}
\renewcommand{\theequation}{S\arabic{equation}}

\bigskip\noindent{\bf\Large Supplementary Materials}

\medskip
\bigskip\noindent{\bf\large Materials and Methods}

\vspace{-3mm}

\paragraph{Determining power flow in each interconnection.}
Each snapshot of each interconnection is modeled as a network of buses connected by transmission lines, 
where a bus represents an end point of a
transmission line or  a coil of a transformer. 
We extracted the following parameters
from the FERC data: the net injected real power at each generator bus, the capacity of each generator, the power demand at each non-generator bus, as well as the parameters of each transmission line and transformer (including their impedance and 
long-term capacity rating).
A transformer is modeled as a transmission line $\ell$ connecting two buses ($i$ and $j$),  where the voltage ratio between these points determines the 
tap
ratio $\tau_\ell$ of this line, and the phase shift of the transformer determines the antisymmetric phase shift matrix $\Delta=(\Delta_{ij})$;
transformers can thus fail in a cascade due to overheating.
The state of the network can be represented by the complex voltage $V_i = |V_i|e^{j\Theta_i}$ at each bus $i$, which in a steady state
is determined by the power flow equation $V_i \cdot I^*_i = S_{i}^{\text{(g)}} - S_{i}^{\text{(d)}}$,  
where $I^*_i$ is the conjugate of the injected complex current,  $S_{i}^{\text{(g)}}=P_{i}^{\text{(g)}}+jQ_{i}^{\text{(g)}}$ is the complex power produced (by the connected generators), 
$S_{i}^{\text{(d)}}=P_{i}^{\text{(d)}}+jQ_{i}^{\text{(d)}}$ is the power consumed (as determined by load demand), and $j := \sqrt{-1}$ is the imaginary unit.  The vector of current injections $I=(I_i)$ is determined by the admittance matrix $Y$ of the network through $I= Y\cdot V$, where $Y$ is a Laplacian matrix whose off-diagonal element $Y_{ij}$ is the negative of the admittance of the lines connecting $i$ and $j$. 
In a steady state,
the total power generated in the network is equal to the total power consumed. Invoking the  DC approximation~(\textit{42}), 
which assumes that the line resistance is negligible and the voltage magnitudes can be approximated as 
$|V_i| \approx 1$ (per unit), 
we linearize the power flow equation to obtain $P^{\text{(g)}}-P^{\text{(d)}}=B\cdot \Theta-P^{\text{shift}}$ for  the real power. Here 
$\Theta=(\Theta_i)$ is the vector of phase angles, $P^{\text{shift}}=(P^{\text{shift}}_i)$ is vector of the phase shifts,   $B$ is a Laplacian matrix with off-diagonal elements 
$B_{ij}=-1/(\tau_\ell x_\ell)$, 
parameter $x_\ell$ is the 
line or transformer
reactance, and the 
tap
ratio 
$\tau_\ell = 1$
for non-transformer lines. The component $P^{\text{shift}}_i$  represents the phase shift at bus $i$ and is determined by the phase shifts (if any) over all the transmission lines connecting to it: $P^{\text{shift}}_i=\sum_{j \in \mathscr{B}(i)} \Delta_{ij}$, where $\mathscr{B}(i)$ is the set of first neighbors of $i$. Then, 
with $\Theta$ determined by the linearized power flow equation, we can approximate the real power flowing on the line $\ell$ from bus $i$ to $j$
by $\dfrac{1}{\tau_\ell x_\ell}(\Theta_i - \Theta_j-\Delta_{ij})$.

\paragraph{Triggering perturbation.}
For a given snapshot of a given interconnection, we model the initial perturbation that triggers a cascade event as the removal of a fixed number $n_\text{t}$ of randomly chosen transmission lines.
For the Texas and Western interconnections, we choose these lines uniformly from all lines present in the snapshot.
For the Eastern interconnection (the largest of the three), we constrain the line selection to one of the six regions 
defined by 
NERC (Fig.~1A and Table~\ref{tab:nerc_region}),
but with the event then evolved through the entire interconnection.
This choice accounts for the intuition that failures in geographical proximity are more likely to trigger large cascades, which we quantitatively 
verified
(Fig.~5) 
and is consistent with the empirical observation that blackout-causing perturbations tend to be 
localized~(\textit{3}).

\paragraph{Modeling cascade dynamics.}
Each iteration of our simulation begins with the removal of a line 
that models an overload failure
(or of the 
$n_{\text{t}}$ 
lines 
chosen as the
initial perturbation for the first iteration).
If the network remains connected after the removal, the redistribution of power flow is calculated by solving the DC power flow equation, 
as described in a section above.
The DC approximation 
offers the computational efficiency that allows for the simulation of
cascading failures in large-scale networks and 
is commonly used
in the engineering 
community~(\textit{13,\,14,\,24,\,25}).
If the grid separates into isolated sub-grids after the line removal, the following procedure is taken to rebalance supply and demand, which allows for the calculation of the redistributed power flow. 
For each sub-grid with unbalanced total power generation and consumption, we first select a generator with the largest capacity as a ``slack bus'' (a generator whose output can be adjusted between zero and its maximum generation output on short time scales) and adjust its output as much as possible within the range allowed;  the slack bus models the role of 
generation
reserves typically designed into real power grids.
If this does not result in balanced supply and demand in a sub-grid,
we uniformly scale down the output of all generators or the consumption (load) at all buses in the sub-grid to achieve a balance, depending on whether the supply is larger or smaller than the demand.
The latter case, in which the total consumption is reduced, models power shedding procedures used by grid operators.
Applying this procedure to all isolated sub-grids, we obtain the redistributed power flow over the network.

Given the redistributed power flow, the next line outage (if any) 
is identified by modeling the heating of line conductors using a 
temperature-evolution 
model~(\textit{12}).
Specifically, 
the temperature of line $\ell$ at time $t$ (measured from the time of flow redistribution within this iteration) is determined by its power flow $P_\ell$ through
\begin{equation}\label{eqn:temp}
T_\ell(t) = e^{-\mu t} \big(T_\ell(0) - T_\text{e} (P_\ell)\big)+T_\text{e}(P_\ell),
\end{equation}
where $T_\text{e}(P_\ell) := \frac{\alpha}{\mu} P_\ell^2 + T_\text{a}$ is the equilibrium temperature [to which $T_\ell(t)$ approaches as $t\to\infty$ according to Eq.~\eqref{eqn:temp}], the constants $\mu$ and $\alpha$ are determined by the properties of the line (assumed to be the same for all $\ell$ for simplicity), and $T_\text{a}$ is the ambient temperature.  %IN WHAT UNITS?
This is a simplified version of the 
temperature-evolution 
model used in 
Ref.~(\textit{12}).
Similar temperature models have also been used 
for
IEEE test systems in recent studies focusing on mitigating 
cascades~(\textit{43}) 
and evaluating 
cascade risk~(\textit{44}).
The capacity of the line $P^{\max}_\ell$ (extracted from the dataset) is associated with the critical temperature $T^\star_{\ell} := T_\text{e}(P^{\max}_\ell)$ above which the line would become overheated.
According to Eq.~\eqref{eqn:temp}, sustained power flow $P_\ell > P^{\max}_\ell$ would bring the line temperature to the critical temperature $T^\star_{\ell}$ at $t = t^\star_{\ell} := - \frac{1}{\mu} \ln \frac{T^\star_{\ell} - T_\text{e} (P_\ell)}{T_\ell(0) - T_\text{e} (P_\ell)}$.
When this 
occurs,
we remove line $\ell$ from the network (primary line failure) to model the action of a protective relay that automatically shuts down the line to prevent permanent damage. 
Note that the critical temperature $T^\star_{\ell}$ serves as a proxy for other failure  
criteria, such as those based on various stability considerations.
The next iteration then begins with the removal of the line with the smallest $t^\star_{\ell}$, followed by the update of the temperature of all the other lines to the value given by Eq.~\eqref{eqn:temp} for $t=\min_\ell t^\star_{\ell}$ and the recalculation of the power flow. 

We repeat this 
process of line removal and power flow redistribution until no line is overheated (i.e., $T_\text{e} (P_\ell) \le T^\star_{\ell}$ or, equivalently, $P_\ell \le P^{\max}_\ell$, for all $\ell$), which generates a (finite) sequence of line failures with time stamps, along with the total power shed.
At the beginning of the cascade event, the temperature of all lines are assumed to be equal to the ambient temperature, $T_\ell(0) = T_\text{a}$ for all $\ell$, which ensures that the failure 
sequence 
and power shed do not depend on $T_\text{a}$, $\alpha$, or $\mu$.

\paragraph{Model validation.}
We 
validated 
the
cascade model against available historical data on cascades 
in the Western interconnection.
We compared the
size of cascades from simulations and from the observed real events 
in terms of two different measures:
the number of primary line failures $N_\text{f}$ and the power shed $P_\text{s}$.
Considering the scarcity of public data on line 
outages (primary failures), 
we 
used
the portion of the Western interconnection represented 
in
the BPA data as a proxy for the entire network. 
Following the criteria in 
Ref.~(\textit{45}), 
we 
identified
the individual cascades 
by grouping the outages based on their temporal proximity, which 
resulted
in $5{,}227$ cascade events from the recorded $8{,}864$ transmission line outages, each triggered by a set of varying number of line failures that 
was also
identified from the data.
To compare with this historical data, we 
simulated
cascades using all available snapshots 
of the Western interconnection.  For this purpose the number of triggers 
$n_{\text{t}}$ 
is chosen randomly following the probability distribution $P(n_\text{t})$ estimated from the 
data.  We 
generated
a total of 
$10^6$ simulated events, 
with the numbers for individual snapshots chosen to be proportional to the weights in Table~\ref{tab:dataset_descrpt}.
Figure~\ref{fig:SI_validate}{A} shows that the sample distribution of cascade sizes generated from these simulations is in good agreement with the distribution from real events when measured in terms of  $N_\text{f}$. 

The NERC data includes
$190$ power outages reported between years 1984 and 2006, among which $93$ cascade events have 
power
shed larger than $300$MW. The dataset formed by these $93$ large cascades is believed to be  complete and reliable given the NERC requirement on the reporting of cascades resulting in  $\ge\!300$MW uncontrolled load.
Using 
both the BPA and NERC data, 
we also 
validated 
the extremal cascade size 
(in terms of 
$N_\text{f}$ and  $P_\text{s}$, respectively)
in our simulations against the historical data (Fig.~\ref{fig:SI_validate}, B and C).

\paragraph{Density-equalizing maps.}
In the diffusion-based algorithm of Ref.~(\textit{41}),
the distribution of 
nodes
is represented by a density function $\rho (\vec r)$ and is then  evolved to a uniform-density equilibrium through a linear diffusion process. 
The displacements of 
nodes
by this diffusion process determine the 
transformation
of the original map into a properly scaled density-equalizing cartogram. 
In our 
implementation used in Fig.~3, 
the density of 
nodes
was calculated with respect to 
a fine-grained  $200 \times 200$ grid of a box including the entire U.S.-South Canada power grid ($24.56^{\circ}$N--$59.15^{\circ}$N, 
$130.35^{\circ}$W--$59.87^{\circ}$W) in the rectangular map projection.
To control the distortion and facilitate interpretation of the result, we 
focused
on the U.S. portion of the grid and 
assigned
a constant nonzero density  everywhere outside the U.S. border.

\paragraph{Identifying mechanisms responsible for primary failures.}
In our cascade model, a given primary failure is caused either by the rerouting of power flow that occurs 
following
the previous primary failure or by the adjustment of generator outputs that may occur when a part of the grid becomes disconnected from the rest (noting that consumption is only adjusted downwards and thus cannot by itself cause overloading).
If the grid 
does not become
disconnected, then 
flow rerouting must be 
responsible for the failure.
If the grid 
becomes
disconnected, 
the mechanism
can be either flow rerouting, generator output adjustment, or both.
In that case, we quantify the extent to which flow rerouting has caused the failure using the following algorithm. 
Let $P_{\ell}(k)$ denote the power flow carried by line $\ell$ at the end of the $k$th iteration (just before the next line removal) and let $\Delta_{\ell}(k) := P_{\ell}(k+1) - P_{\ell}(k)$ be the total flow change on line $\ell$ from the $k$th to the $(k+1)$th iteration.
Then, the
fraction $f_{\ell i}$ of the flow $P_{\ell}(k)$ that is supplied by generator $i$ can be defined and computed using a flow tracing algorithm based on a proportional sharing 
principle~(\textit{46}).
This fraction can be used to determine the 
amount of flow change $\Delta^{(\text{g})}_{\ell i}(k)$ 
that can be attributed to the adjustment of generator $i$ as 
$\Delta^{(\text{g})}_{\ell i}(k) := f_{\ell i} \Delta P^{(\text{g})}_i(k)$, 
where 
$\Delta P^{(\text{g})}_i(k)$ 
denotes the change in 
the output of generator $i$
in this iteration.
The total amount of flow change caused by the generation adjustments is then 
$\Delta^{(\text{g})}_{\ell}(k) := \sum_i \Delta^{(\text{g})}_{\ell i}(k)$.
Thus, 
$\Delta^{(\text{g})}_{\ell}(k)$ 
quantifies the degree to which generation changes have contributed to the failure, if line $\ell$ fails at the beginning of the $(k+1)$th iteration.
The contribution of flow rerouting to the failure 
is then
measured by 
$\Delta^{(\text{r})}_{\ell}(k) := \Delta_{\ell}(k) - \Delta^{(\text{g})}_{\ell}(k)$.
If the grid were not disconnected in that iteration, we would have 
$\Delta^{(\text{g})}_{\ell}(k) = 0$ and $\Delta^{(\text{r})}_{\ell}(k) = \Delta_{\ell}(k)$, 
since the whole change would have been due to rerouting.
To account for cases in which a line continues to be overloaded for multiple iterations before experiencing a primary failure, we keep track of running totals 
$\widetilde{\Delta}^{(\text{r})}_{\ell} := \sum_k \Delta^{(\text{r})}_{\ell}(k)$, $\widetilde{\Delta}^{(\text{g})}_{\ell} := \sum_k \Delta^{(\text{g})}_{\ell}(k)$, 
and $\widetilde{\Delta}_{\ell} := \sum_k \Delta_{\ell}(k)$ for each $
\ell$, where the sums are taken over the most recent set of consecutive iterations in which the line has been 
overloaded.
If, at the time of a primary failure, 
$\widetilde{\Delta}^{(\text{r})}_{\ell}$ and $\widetilde{\Delta}_{\ell}$ have
the same sign but $\widetilde{\Delta}^{(\text{g})}_{\ell}$ has the opposite sign
(or $\widetilde{\Delta}^{(\text{g})}_{\ell}=0$), 
then flow rerouting must be 
solely
responsible for the failure (as 
$\widetilde{\Delta}^{(\text{g})}_{\ell}$ 
can only help prevent overloading in that case).
Using the 2010 spring peak snapshot of the Texas grid as a representative example, we 
counted 
the number of 
primary failures
for which this 
was
the case 
among
all primary failures observed in $5{,}000$ simulated cascade events.
We 
found
this number to be $2{,}222$ out of $2{,}257$ observed primary failures, indicating that 
rerouting of power flow 
was solely responsible for
over $98$\% of these primary failures.

\paragraph{Distances between transmission lines.} 
To define the topological and geographical distances between two lines, we start from the corresponding notions of distance between two buses. The topological distance $d_{ij}$ is defined as the number of lines along the shortest path between buses $i$ and $j$ in the network. The geographical distance $g_{ij}$ is defined as the arc length between buses $i$ and $j$ along the Earth's great circle. 
The topological distance $d_{\ell \ell'}$ between line $\ell$ (connecting bus $i$ to $j$) and line $\ell'$ (connecting bus $i'$ to $j'$) is defined as the minimum of the set $\{ d_{ii'},d_{ij'},d_{ji'},d_{jj'} \}$. 
The distance between line $\ell$ and a set of lines $S$
is defined as $\min_{\ell' \in S} \{d_{\ell \ell'}\}$
and the distance between two sets, $S$ and $S'$, 
is defined as $\min_{\ell \in S, \ell' \in S'} \{d_{\ell \ell'}\}$.
The geographical distances are defined similarly, with each $d$ replaced by $g$ in the notations. 
When calculating these distances in our characterization of cascading failures, we normalize them by the average pairwise distance between any two  lines in the corresponding NERC region  where the cascade events are 
triggered.

\clearpage
\hoffset 0cm
\textwidth 16cm

%:Supplementary Figures
\clearpage
\bigskip
\noindent{\bf\large Supplementary Figures}

\newcounter{sfigure}
\renewcommand{\figurename}{Fig.}
\renewcommand{\thefigure}{S\arabic{sfigure}}

\vspace{5mm}

\addtocounter{sfigure}{1} 
\begin{figure*}[h!] 
\begin{center}
\includegraphics[width=1.0\textwidth]{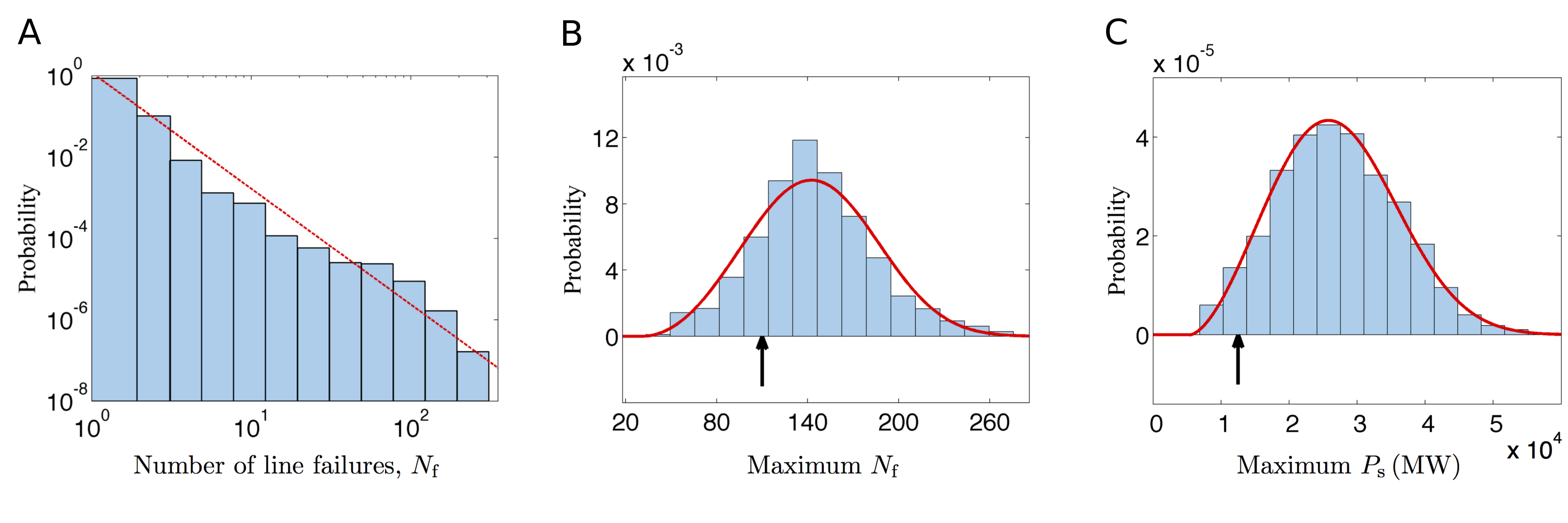}
\caption{\label{fig:SI_validate}
\textbf{Validation of the cascade model against historical data for the Western interconnection}. 
The cascade size from simulations and from real events are compared by the number of (primary) line failures $N_\text{f}$ for the BPA data comprised of a total of $5{,}227$ cascades, and by the amount of power shed  $P_\text{s}$ for the NERC data comprised of $93$ large cascades ($\ge\! 300$MW).
(\textbf{A}) Histogram of the cascade size ($N_\text{f}$) from simulations  in which the number of triggering line failures follows the empirical distribution derived from the BPA data. 
The simulation results are in agreement with the linear regression (dashed line) of the 
BPA data compiled in 
Ref.~(\textit{45}).
(\textbf{B})~Histogram of the largest cascade size in a set of  $5{,}227$ samples of $N_\text{f}$ drawn from
the probability distribution of cascade size computed from the simulations, 
for $5{,}000$ independent realizations of this set (blue). 
Due to the rareness of large cascade events and the corresponding under-sampling in simulations, we 
used
the kernel density estimation [for 
a Gaussian kernel with its width chosen according to the Silverman's rule of 
thumb~(\textit{48})] 
to approximate the $0.1\%$ tail of the distribution.
The $p$-value of the largest $N_\text{f}$ from the BPA historical data (black arrow) 
was found to be 
$0.42$ under the hypothesis that the observed largest  $N_\text{f}$ follows the generalized extreme value 
distribution~(\textit{49}) 
fitted to the histogram (red curve).
(\textbf{C})~Same as (B) but for $P_\text{s}$, using sets of $93$ samples drawn from
the probability distribution computed from simulations, considering only the large 
simulated
cascades. 
Each maximum $P_\text{s}$ 
was 
adjusted by a factor $(0.98)^y$ to incorporate 
approximately $2\%$ annual increase of the power 
demand, 
where $y$ is an integer randomly chosen from $2$ to $14$ to trace
back from
the year of the simulated data (2008)
to years
of historical data 
($1984$--$2006$).
We found the
$p$-value of the largest $P_\text{s}$ from the NERC historical data (black arrow) 
to be
$0.07$ under 
 the hypothesis that the observed largest $P_\text{s}$ follows the generalized extreme value distribution 
fitted to the histogram (red curve).
 In both (B) and (C), we cannot reject the null hypothesis that the real largest cascade size follows the fitted distribution at a significance level of $5\%$. 
}
\end{center}
\end{figure*}

\addtocounter{sfigure}{1} 
\begin{figure*}[h] 
\begin{center}
\includegraphics[width=0.6\textwidth]{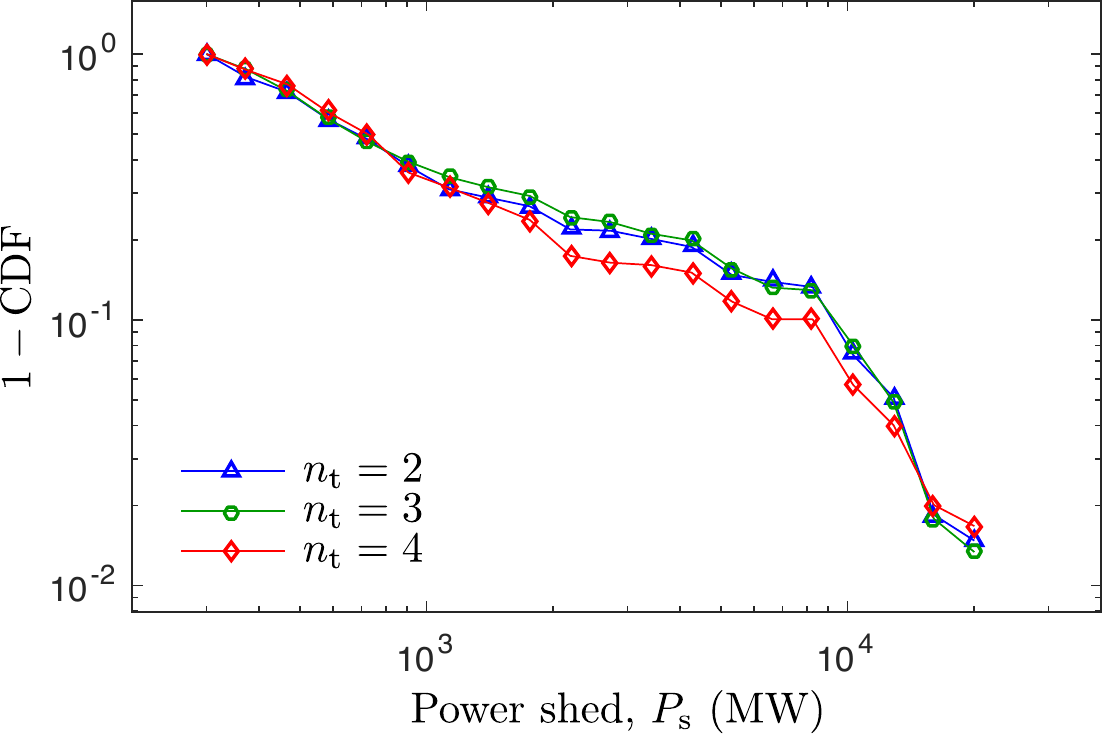}\rule{10mm}{0mm}\\[2mm]
\caption{\label{fig:SI_validate_difftrig}
\textbf{Robustness of cascade size distribution with respect to the number of triggering line failures}. 
Cumulative distribution functions (CDF) of the cascade size $P_\text{s}$ for large 
simulated
cascades  ($\ge\! 300$MW) triggered by disabling $n_\text{t}=2$ (blue), 
$n_\text{t}=3$ (green), and $n_\text{t}=4$ (red) power lines
in the Western interconnection. 
For each $n_\text{t}$, the CDF was calculated by combining the estimated distribution of $P_\text{s}$ for all snapshots using the weights given in Table~\ref{tab:dataset_descrpt}.
The Kolmogorov-Smirnov 
test~(\textit{47}) 
cannot reject the hypothesis that the size of cascades triggered by $n_\text{t}=3$ line failures has the same underlying distribution 
as of those triggered by $4$ line failures  ($p$-value = $0.22$) or $2$ line failures  ($p$-value = $0.08$).
This supports the conclusion that the distribution does not depend sensitively on the size of the perturbation, even though the size of the largest cascade might.}
\end{center}
\end{figure*}

\addtocounter{sfigure}{1}
\begin{figure*}[p]
\vspace{-15mm}
\begin{center}
\includegraphics[width=.92\textwidth,scale=0.3]{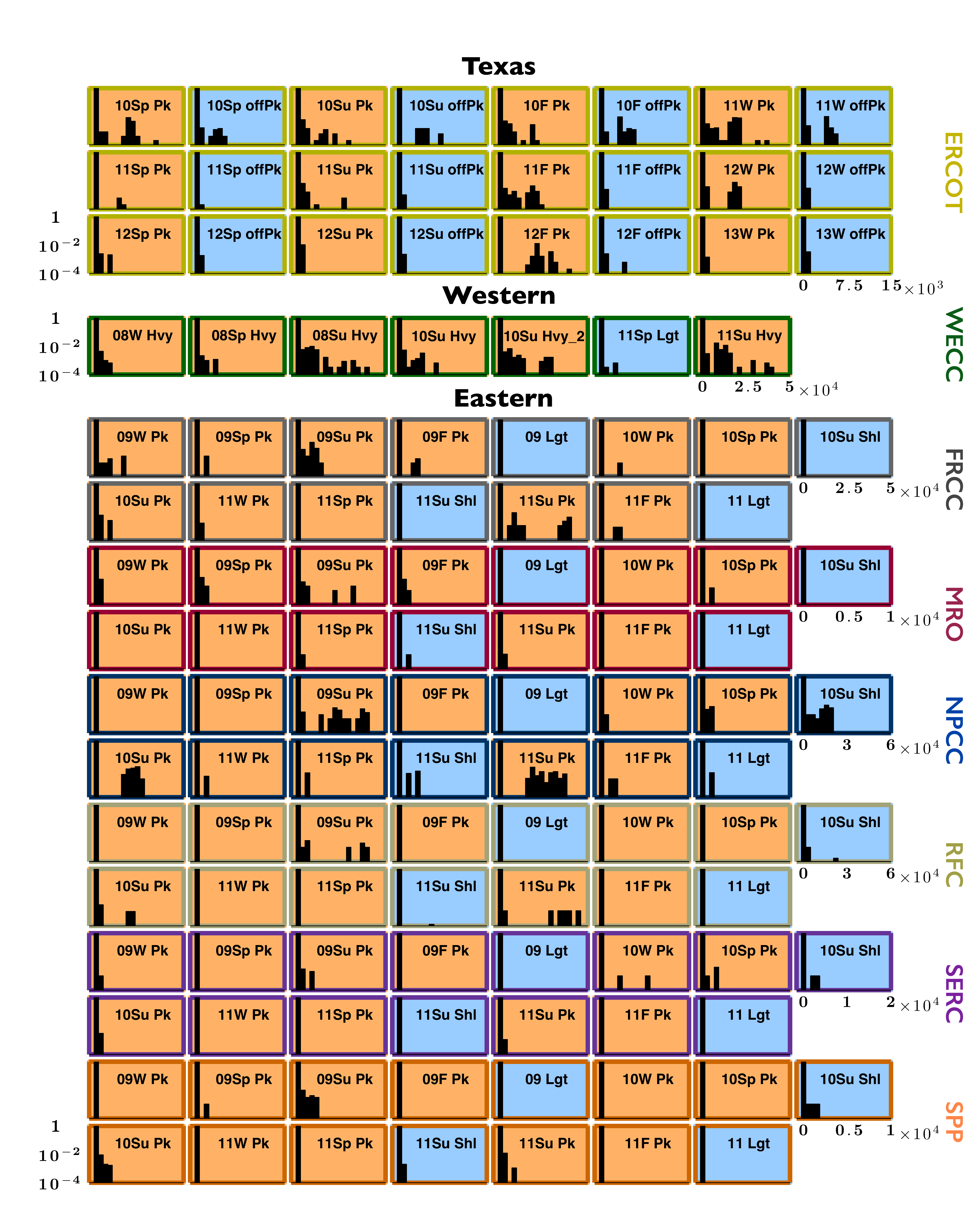} \\
\vspace{-2mm}
\caption{\label{fig:cascade_hist_ps}
\textbf{Distribution of cascade size from simulated events broken down by region and snapshot.} Histograms of the relative frequencies of cascade size $P_\text{s}$ (measured in MW) for events triggered in different NERC regions (marked by the color of the box border), where each panel represents a different snapshot of the power grid
for the given region. The acronyms for the snapshots and the abbreviations for the triggering regions are defined in Tables~\ref{tab:dataset_descrpt} and \ref{tab:nerc_region}, respectively.
 The distributions have the same horizontal linear scale within each region and the same vertical logarithmic scale for all regions. The background color indicates the level of power demand for each snapshot, with orange for high demand and blue for moderate and low demand.}
\end{center}
\end{figure*}

\addtocounter{sfigure}{1}
\begin{figure*}[h!]
\begin{center}
\includegraphics[width=.92\textwidth,scale=0.3]{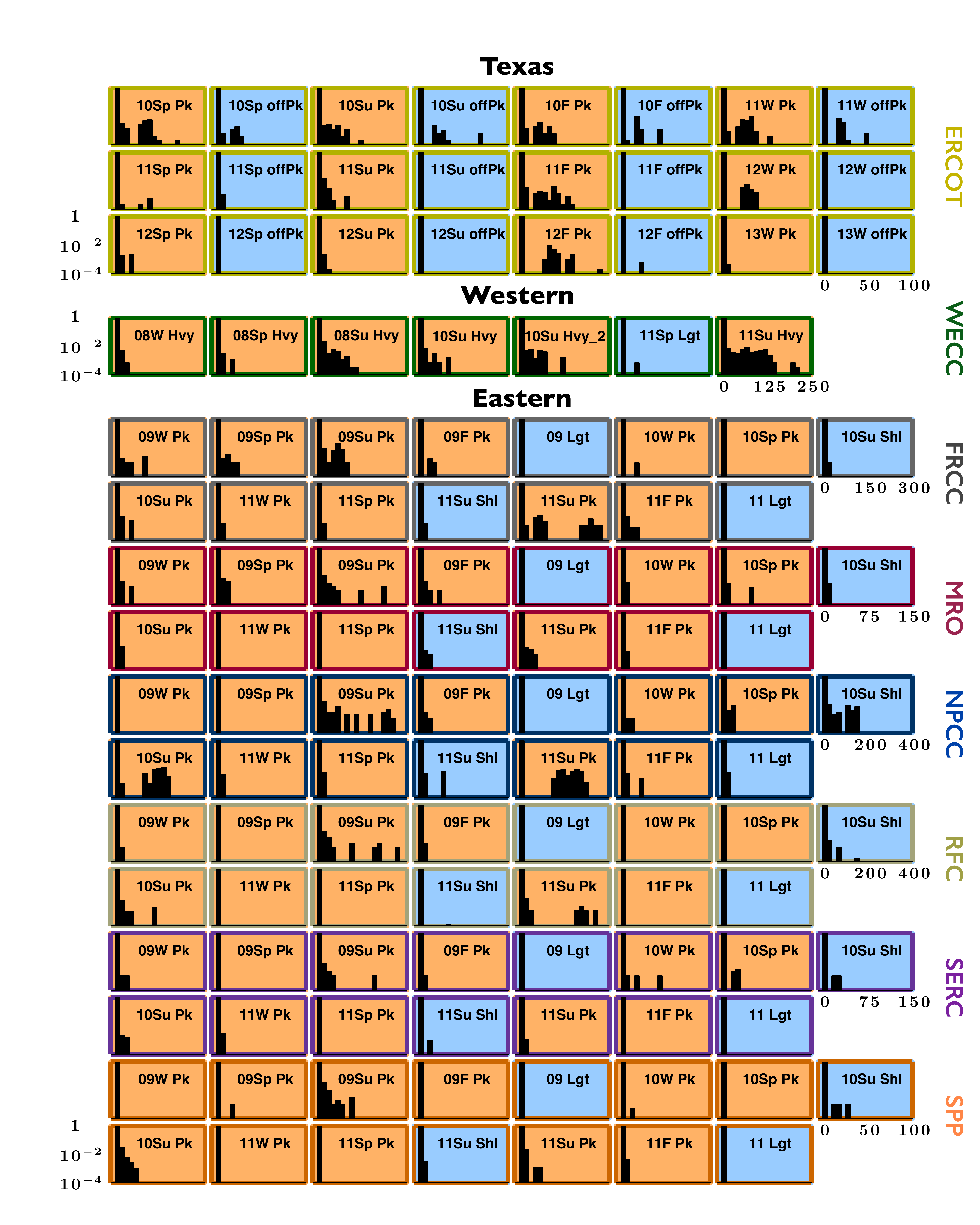}\\
\vspace{-2mm}
\caption{\label{fig:cascade_hist_nf}
\textbf{Same as Fig.~\ref{fig:cascade_hist_ps} but now for cascade size measured by the number of (primary) line failures, $\boldsymbol{N_\text{f}}$.}}
\end{center}
\end{figure*}

\addtocounter{sfigure}{1}
\begin{figure*}[h!]
\begin{center}
\includegraphics[width=.85\textwidth]{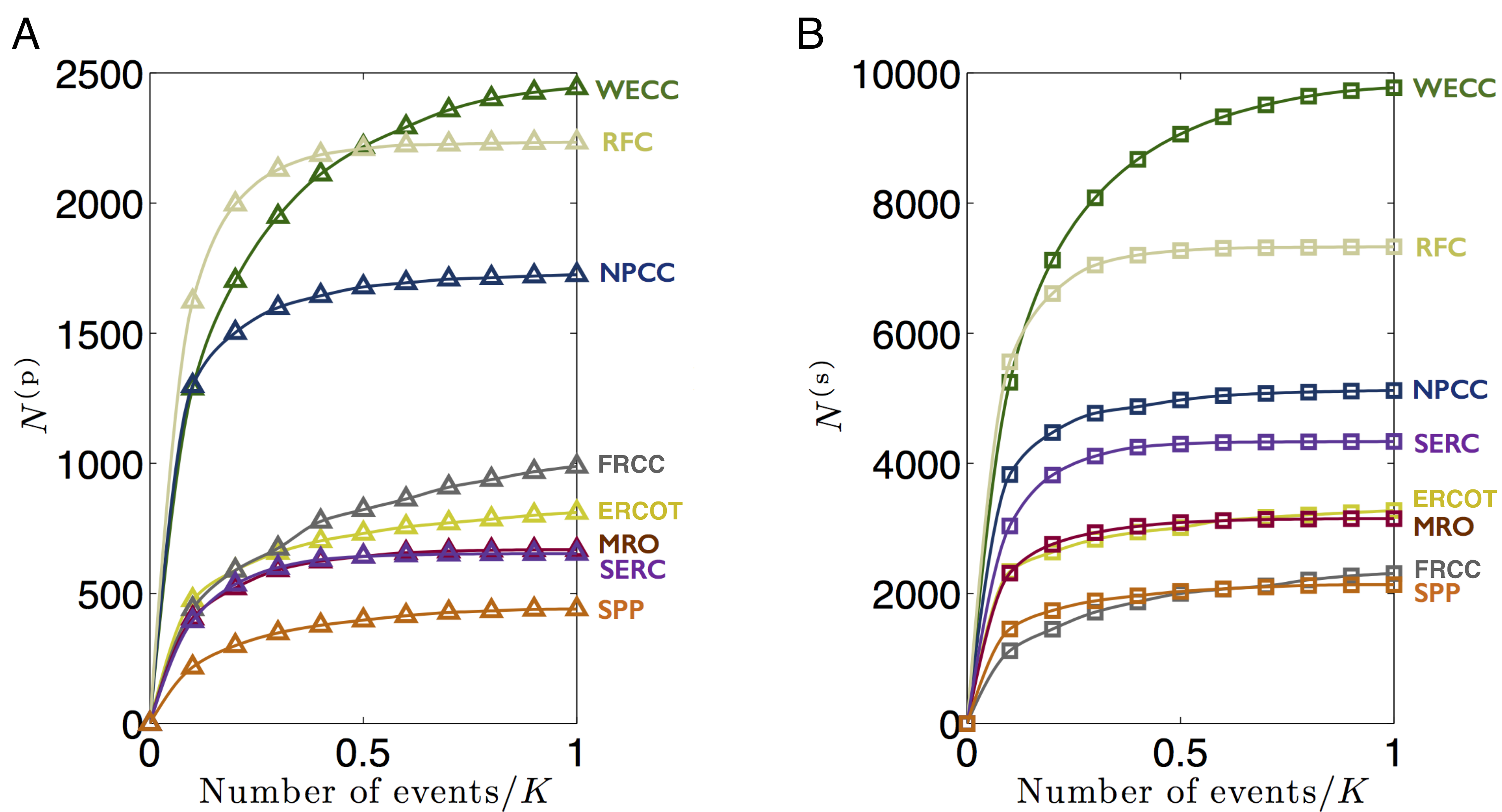}\\[2mm]
\caption{\label{fig:NpNs}
\textbf{Dependence of the number of failing links on the number of simulated events.}
(\textbf{A} and \textbf{B}) Number of links $N^{\text{(p)}}$ that experience primary failures (A) and number of links $N^{\text{(s)}}$ that 
experience secondary failures  (B) versus the (normalized) number of simulated events 
sampled randomly from all $K$ events.
Each symbol is colored by the region of the 
triggering line failures and represents an average over $100$ independent 
samples.
Both $N^{\text{(p)}}$ and $N^{\text{(s)}}$ grow increasingly slower and appear to saturate as the sample size increases, justifying our choices of $K$ given in Table~\ref{tab:dataset_descrpt}.}
\end{center}
\end{figure*}

\addtocounter{sfigure}{1} 
\begin{figure*}[h!]
\vspace{5mm}
\begin{center}
\includegraphics[width=5.5in]{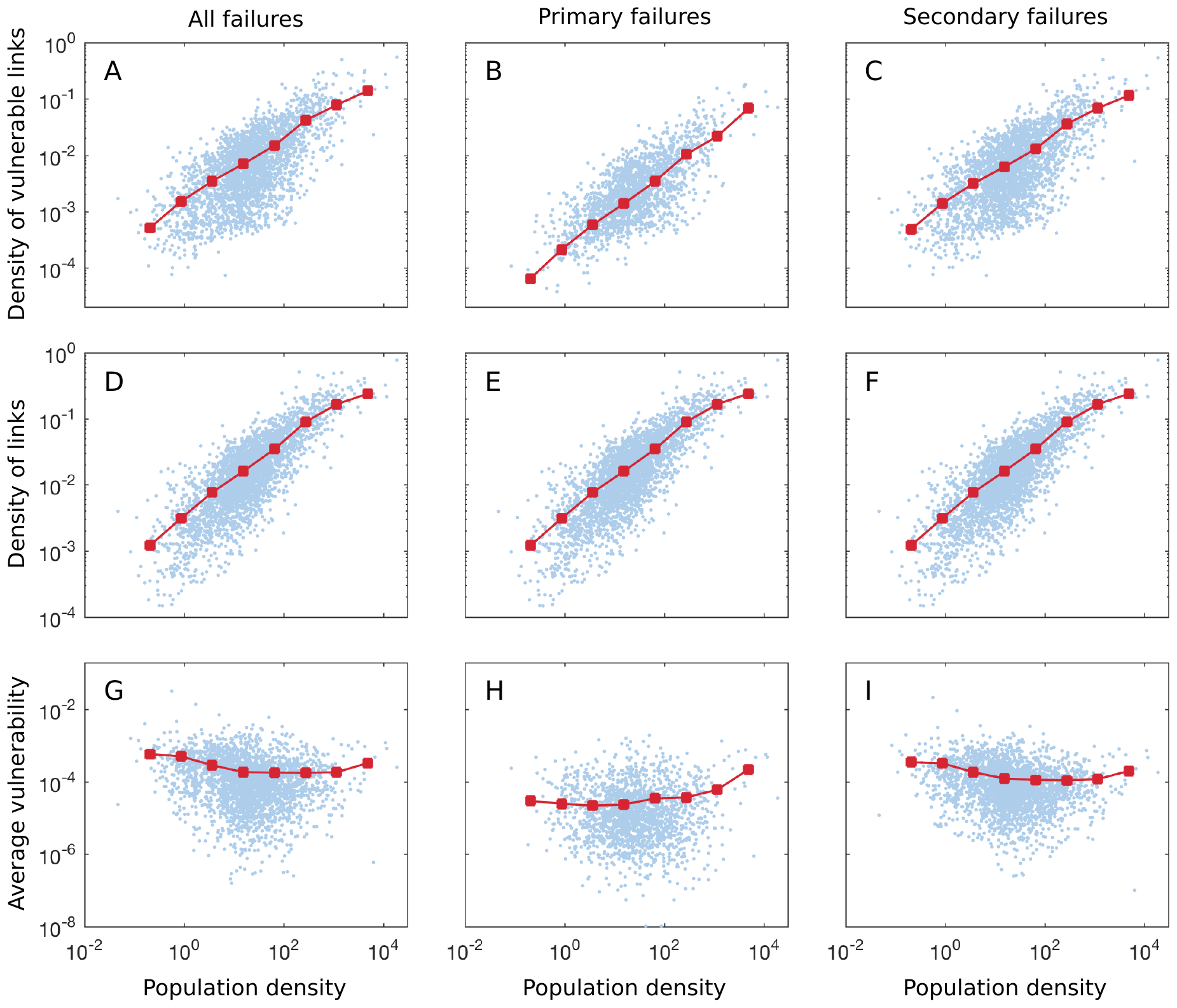}
\caption{\label{fig:density_vs_vulnerability}
\textbf{Correlation between link vulnerability and population density}. 
(\textbf{A}) Density of links with nonzero {\sc a}-vulnerability $\langle p^{\text{\color{white}p}}_\ell \rangle$ in the contiguous U.S.\ as a function of population density.
For each county, the population density (in km$^{-2}$) was determined from the 2010 census data and the geographic coordinates of the county's 
boundaries.
The density of vulnerable links for a county was computed by counting the number of links with $\langle p^{\text{\color{white}p}}_\ell \rangle > 0$ that are connected to nodes within its boundaries, divided by the area of the county (in km$^2$).
The blue scatter points correspond to individual counties, whereas the red curve indicates the average over multiple counties computed by logarithmic binning with respect to population density.
(\textbf{B}~and \textbf{C}) The same as in (A) but shown separately for the vulnerability to primary failures $\langle p^{\text{(p)}}_\ell \rangle $ (B) and the vulnerability to secondary failures $\langle p^{\text{(s)}}_\ell \rangle $ (C).
(\textbf{D} to \textbf{F}) Density of all links (vulnerable or not) as a function of population density, showing positive correlation similar to 
that observed in (A to C).
(\textbf{G}~to~\textbf{I}) 
Average {\sc a}-vulnerability (over all links connected to nodes in a given county) 
as a function of 
population density, showing no clear correlation.
}
\end{center}
\end{figure*}

\addtocounter{sfigure}{1} 
\begin{figure*}[h!]
\begin{center}
\includegraphics[width=1.0\textwidth]{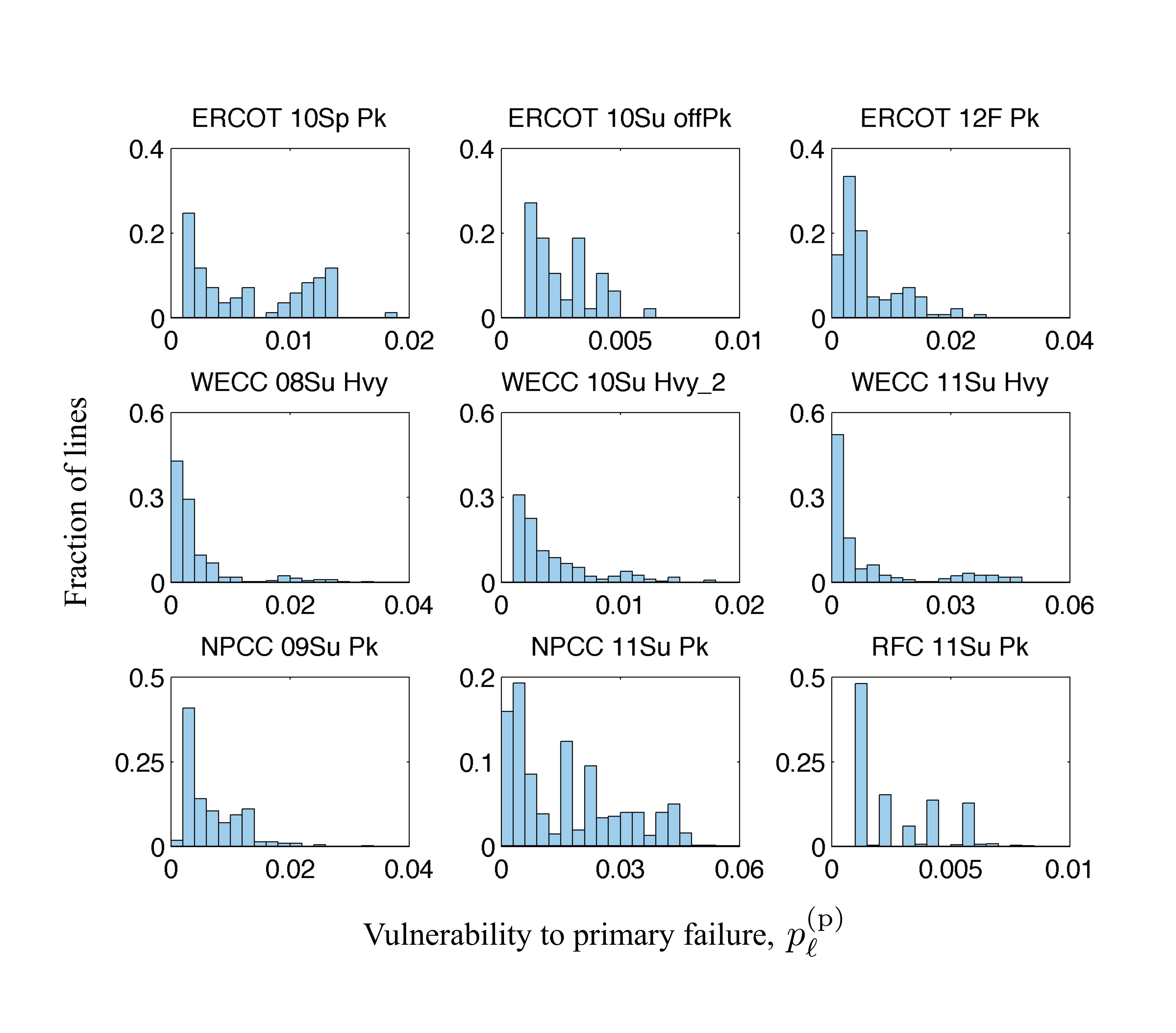}
\vspace{-10mm}
\caption{\label{fig:SI_corr}
\textbf{Vulnerability distribution for individual snapshots.} Histograms of the relative frequencies of line vulnerability to primary failure $p_\ell^{\text{(p)}}$ 
within the vulnerable set, for the top three snapshots in each interconnection when ranked by the size $P_\text{s}$ of the largest 
cascade. 
The abbreviations for the snapshots and
the triggering regions are given in Tables~\ref{tab:dataset_descrpt} and \ref{tab:nerc_region}, respectively.}
\end{center}
\end{figure*}

%:Supplementary Tables

\newcounter{supptable}
\renewcommand{\tablename}{Table}
\renewcommand{\thetable}{S\arabic{supptable}}

\clearpage
\hoffset -0.75cm
\oddsidemargin 0cm
\textwidth 18.3cm 

\addtocounter{supptable}{1}

\begin{table*}[h!]
\vspace{-10mm}
\noindent{\bf\large Supplementary Tables}\\[-5mm]
\begin{center}
\caption{\makebox[18.3cm][l]{\small Case description of the available snapshots of the U.S.-South Canada power grid and simulation details.}}
\vspace{2mm}
\label{tab:dataset_descrpt}
\fontsize{9}{1.5}\selectfont
\begin{tabular}{lllrrrrrr}
\toprule
Interconnection & Snapshot & Acronym & Buses &  Generators & \quad \quad \quad Lines & Load (MW) & $K$ & $w_c$\\
\midrule[0.3pt]						  
\multirow{24}{*}{Texas} 		  & 2010 spring peak & 10Sp Pk & $5{,}960$ & $651$ &$7{,}334$& $61{,}397$& $5{,}000$ & $1/24$\\
						  & 2010 spring off-peak & 10Sp offPk & $5{,}959$ & $651$ &$7{,}331$& $37{,}737$& $5{,}000$ & $1/24$\\
						  & 2010 summer peak & 10Su Pk & $6{,}005$ & $658$ &$7{,}395$& $72{,}095$& $5{,}000$ & $1/24$\\
						  & 2010 summer off-peak & 10Su offPk & $6{,}005$ & $660$ &$7{,}395$& $42{,}239$& $5{,}000$ & $1/24$\\
						  & 2010 fall peak & 10F Pk & $6{,}010$ & $668$ &$7{,}401$& $57{,}663$ & $5{,}000$ & $1/24$\\
						  & 2010 fall off-peak & 10F offPk & $6{,}010$ & $663$ &$7{,}401$& $35{,}342$& $5{,}000$ & $1/24$\\
						  & 2011 winter peak & 11W Pk & $6{,}012$ & $665$ &$7{,}407$& $57{,}252$& $5{,}000$ & $1/24$\\
						  & 2011 winter off-peak & 11W offPk & $6{,}012$& $665$  &$7{,}407$& $38{,}045$& $5{,}000$ & $1/24$\\
						  
						  & 2011 spring peak & 11Sp Pk & $6{,}034$ & $677$  &$7{,}451$ & $56{,}597$& $5{,}000$ & $1/24$\\
						  & 2011 spring off-peak & 11Sp offPk & $6{,}034$ & $679$ &$7{,}451$& $34{,}698$& $5{,}000$ & $1/24$\\
						  & 2011 summer peak & 11Su Pk  & $6{,}057$ & $682$ &$7{,}493$& $70{,}313$& $5{,}000$ & $1/24$\\
						  & 2011 summer off-peak & 11Su offPk & $6{,}057$ & $684$ &$7{,}493$& $40{,}028$& $5{,}000$ & $1/24$\\
						  & 2011 fall peak & 11F Pk & $6{,}062$ & $690$ &$7{,}489$& $55{,}625$& $5{,}000$ & $1/24$\\
						  & 2011 fall off-peak & 11F offPk & $6{,}063$ & $691$ &$7{,}491$& $34{,}148$& $5{,}000$ & $1/24$\\
						  & 2012 winter peak & 12W Pk & $6{,}066$ & $691$ &$7{,}506$& $55{,}951$& $5{,}000$ & $1/24$\\
						  & 2012 winter off-peak & 12W offPk & $6{,}066$ & $691$ &$7{,}506$& $37{,}305$& $5{,}000$ & $1/24$\\
						  
						  & 2012 spring peak & 12Sp Pk & $6{,}405$ & $646$ &$8{,}006$& $56{,}926$& $5{,}000$ & $1/24$\\
						  & 2012 spring off-peak & 12Sp offPk & $6{,}405$ & $646$ &$8{,}006$& $34{,}778$& $5{,}000$ & $1/24$\\
						  & 2012 summer peak & 12Su Pk & $6{,}428$ & $648$ &$8{,}039$& $72{,}442$& $5{,}000$ & $1/24$\\
						  & 2012 summer off-peak & 12Su offPk & $6{,}428$ & $648$ &$8{,}039$& $40{,}934$& $5{,}000$ & $1/24$\\
						  & 2012 fall peak & 12F Pk & $6{,}432$ & $649$ &$8{,}043$& $57{,}441$& $5{,}000$ & $1/24$\\
						  & 2012 fall off-peak & 12F offPk & $6{,}432$ & $649$ &$8{,}043$& $34{,}810$& $5{,}000$ & $1/24$\\
						  & 2013 winter peak & 13W Pk & $6{,}459$ & $667$ &$8{,}075$& $60{,}209$& $5{,}000$ & $1/24$\\
						  & 2013 winter off-peak & 13W offPk & $6{,}459$ & $667$ &$8{,}075$& $39{,}345$& $5{,}000$ & $1/24$\\

\midrule[0.3pt]

\multirow{7}{*}{Western}    	  & 2008 heavy winter & 08W Hvy & $15{,}387$ & $2{,}997$ &$19{,}687$& $128{,}331$& $6{,}000$ & $3/24$ \\
					          & 2008 heavy spring & 08Sp Hvy & $15{,}439$ & $3{,}017$ &$19{,}752$& $113{,}824$& $6{,}000$ & $6/24$\\
						  & 2008 heavy summer & 08Su Hvy & $15{,}722$ & $3{,}065$ &$20{,}037$& $159{,}890$ & $6{,}000$ & $1/24$\\						
						  & 2010 heavy summer & 10Su Hvy & $16{,}932$ & $3{,}189$ &$20{,}476$& $163{,}174$& $6{,}000$ & $(1/2){\cdot}1/24$ \\
						  & 2010 heavy summer\hspace{-0.5pt}\_\hspace{0.3pt}2 & 10Su Hvy\hspace{-1pt}\_\hspace{0.3pt}2 & $16{,}796$ & $3{,}292$ &$21{,}269$& $168{,}091$& $6{,}000$ & $(1/2){\cdot}1/24$\\
						  & 2011 light spring & 11Sp Lgt & $16{,}082$ & $3{,}181$ &$20{,}711$& $98{,}415$& $6{,}000$ & $12/24$\\
						  & 2011 heavy summer &11Su Hvy & $17{,}524$ & $3{,}467$ &$22{,}510$& $194{,}983$& $6{,}000$& $1/24$\\
						  						  
\midrule[0.3pt]		  
						  
\multirow{15}{*}{Eastern} 	          & 2009 winter peak & 09W Pk & $53{,}887$ & $7{,}683$ &$69{,}431$& $548{,}727$& $23{,}000$ & $1/24$\\						  
						  & 2009 spring peak & 09Sp Pk & $53{,}649$ & $7{,}645$ &$69{,}179$& $460{,}594$ & $23{,}000$ & $1/24$\\
						  & 2009 summer peak & 09Su Pk & $54{,}025$ & $7{,}614$ &$69{,}697$& $641{,}388$& $23{,}000$ & $1/24$\\
						  & 2009 fall peak & 09F Pk & $53{,}866$ & $7{,}676$ &$69{,}495$& $474{,}961$ & $23{,}000$ & $(3/2){\cdot} 1/24$\\
						  & 2009 light load & 09 Lgt & $53{,}651$ & $7{,}584$ &$69{,}182$& $286{,}672$& $23{,}000$& $6/24$\\
						  & 2010 winter peak &10W Pk & $56{,}714$ & $7{,}932$ &$72{,}935$& $538{,}747$& $23{,}000$ & $1/24$\\
						  & 2010 spring peak & 10Sp Pk & $56{,}422$ & $7{,}847$ &$72{,}519$& $452{,}769$ & $23{,}000$ & $1/24$\\
						  & 2010 summer shoulder & 10Su Shl & $56{,}616$ & $7{,}804$ &$72{,}780$ & $492{,}851$ & $23{,}000$ & $(1/2){\cdot}1/24$\\
						  & 2010 summer peak & 10S Pk & $56{,}642$ & $7{,}805$ &$72{,}811$& $622{,}180$ & $23{,}000$ & $(1/2){\cdot}1/24$ \\
						  & 2011 winter peak &11W Pk & $60{,}073$ & $8{,}137$ &$76{,}942$& $537{,}272$ & $23{,}000$ & $1/24$\\
						  & 2011 spring peak &11Sp Pk & $59{,}741$ & $8{,}044$ &$76{,}487$& $456{,}678$ & $23{,}000$ & $1/24$\\
						  & 2011 summer shoulder &11Su Shl & $59{,}943$ & $8{,}027$ &$76{,}791$& $490{,}586$ & $23{,}000$ & $(1/2){\cdot}1/24$\\
						  & 2011 summer peak &11Su Pk & $59{,}948$ & $8{,}043$ &$76{,}798$& $616{,}200$ & $23{,}000$ & $(1/2){\cdot}1/24$\\				  
						  & 2011 fall peak &11F Pk & $59{,}966$ & $8{,}089$ &$76{,}798$& $467{,}659$ & $23{,}000$ & $(3/2){\cdot}1/24$\\
						  & 2011 light load &11 Lgt & $59{,}669$ & $7{,}993$ &$76{,}395$& $269{,}961$ & $23{,}000$ & $6/24$\\

\bottomrule
\end{tabular}
\end{center}
\makebox[16cm][l]{\small
\begin{minipage}{18.3cm}
The description of each snapshot follows the terminology in the FERC Form 715 and is abbreviated to facilitate referencing in 
this
article.
The number of buses, number of generators, number of transmission lines, and the amount of load vary across different snapshots. Unless noted otherwise, our simulation results  are based on $K$ simulated events in which the network is perturbed by the removal of $n_\text{t}=3$ randomly selected transmission lines within the same region. For the Eastern interconnection, which consists of six  regions 
(Fig.~1A and Table~\ref{tab:nerc_region})
the $23{,}000$ simulations are distributed approximately proportionally to the number of transmission lines in each region:  FRCC ($1{,}200$), MRO ($3{,}400$), NPCC ($4{,}000$), RFC ($6{,}400$), SERC ($6{,}000$), and SPP ($2{,}000$). The last column lists the relative weights assigned to each snapshot in our calculations of averages, chosen to give comparable weights across the different seasons as well as to  high and low power demand conditions.
\end{minipage}}
\end{table*}

\clearpage
\hoffset -1cm
\textwidth 18cm 

\addtocounter{supptable}{1}

\begin{table*}[h!]
\begin{center}
\caption{Names, abbreviations, and basic properties of the NERC  reliability regions.} 
\label{tab:nerc_region}
\resizebox{\textwidth}{!}{
\begin{tabular}{llccc}
\toprule
Interconnection &  Region & Buses & Lines & Load (MW)\\  
\midrule
Texas & Electric Reliability Council of Texas (ERCOT) & $\enskip5{,}959$ -- $\enskip6{,}459$ & $\enskip7{,}331$ -- $\enskip8{,}075$ 
& $34{,}148$ -- $\enskip72{,}442$\\
\midrule[0.3pt]
Western & Western Electricity Coordinating Council (WECC) & $15{,}387$ -- $17{,}524$ & $19{,}687$ -- $22{,}510$  
&$98{,}415$ -- $194{,}983$\\
\midrule[0.3pt]
\multirow{6}{*}{Eastern} & Florida Reliability Coordinating Council (FRCC) & $\enskip2{,}945$ -- $\enskip3{,}022$ & $\enskip3{,}795$ -- $\enskip3{,}914$  
& $17{,}845$ -- $\enskip49{,}608$\\
&Midwest Reliability Organization (MRO) & $\enskip7{,}605$ -- $\enskip8{,}340$ & $\enskip9{,}370$ -- $10{,}297$  
& $24{,}601$ -- $\enskip49{,}275$\\
&Northeast Power Coordinating Council (NPCC) & $\enskip9{,}381$ -- $10{,}784$ & $12{,}609$ -- $14{,}054$ 
&$42{,}398$ -- $\enskip92{,}524$\\
&ReliabilityFirst Corporation (RFC) 
& $14{,}896$ -- $17{,}497$ & $19{,}753$ -- $23{,}152$  
&$81{,}970$ -- $201{,}736$\\ 
&SERC Reliability Corporation (SERC)  
& $13{,}751$ -- $15{,}320$ & $17{,}637$ -- $19{,}444$ 
 &$84{,}676$ -- $211{,}561$\\
&Southwest Power Pool Regional Entity (SPP) 
& $\enskip4{,}969$ -- $\enskip5{,}161$ & $\enskip5{,}908$ -- $\enskip6{,}126$ 
 & $14{,}203$ -- $\enskip38{,}894$\\
\bottomrule
\multicolumn{5}{l}{
For each property, we show the minimum and the maximum 
over all snapshots in our dataset (listed in Table~\ref{tab:dataset_descrpt}).
}
\end{tabular}
}
\end{center}
\end{table*}

\clearpage
\hoffset 0cm
\oddsidemargin 0.2cm
\textwidth 16cm 

\addtocounter{supptable}{1}
\begin{table*}[h!]
\begin{center}
\caption{Number of links experiencing primary and secondary failures.}
\label{tab:NpNs}
\begin{tabular}{lrrrrr}
\toprule
\vspace{1mm}
Interconnection 
& Nodes & Links & $N^\text{(p)}$ & $N^{\text{(s)}}$ &$\frac{N^\text{(s)}}{N^{\text{(p)}}}$ \\
\midrule
Texas  & $3{,}275$  & $6{,}664$ &$811$ & $3{,}274$ & $4.04$\\
Western & $9{,}658$ & $21{,}589$ & $2{,}443$  &  $9{,}773$ & $4.00$\\
Eastern & $27{,}822$ & $54{,}846$ & $5{,}726$ & $20{,}798$ & $3.63$\\
\quad FRCC &$1{,}598$&$3{,}188$& $988$ & $2{,}306$ & $2.33$\\								
\quad MRO & $3{,}972$& $6{,}952$& $668$ & $3{,}150$ & $4.72$\\
\quad NPCC& $3{,}253$ &$7{,}758$& $1{,}725$ & $5{,}120$ & $2.97$\\
\quad RFC  &$8{,}247$& $17{,}674$ & $2{,}234$ & $7{,}328$ & $3.28$\\
\quad SERC &$8{,}619$ & $17{,}969$ & $653$ & $4{,}335$ & $6.64$\\
\quad SPP &$2{,}501$ & $5{,}312$ & $440$ & $2{,}136$ & $4.85$\\
\midrule
Total & $40{,}755$ & $83{,}099$ & $8{,}980$ & $33{,}845$ & $3.77$\\
\bottomrule
\end{tabular}
\end{center}
\vspace{2mm}
{\small
We show the number of links $N^{\text{(p)}}$ $\bigl(N^{\text{(s)}}\bigr)$ whose {\sc a}-vulnerability to primary (secondary) failures was nonzero in our simulations for each NERC region in which cascade events were triggered, along with the total for the entire network.
Among the $83{,}099$ links in the network, the number of links that experienced primary failures was $8{,}980$ ($10.8\%$). 
For the Eastern interconnection, there are overlaps between the failed links in cascades triggered from different regions,  which is the reason why the number of links that fail in the entire interconnection is smaller than the corresponding number  summed over the 
six
regions. 
For comparison, the second and third columns list the total number of 
nodes
and links, 
respectively.  For the Eastern interconnection, 
these totals are also smaller than the corresponding sums over the individual regions,  in this case because  some small portions of the grid belong to different NERC regions at different times.}
\end{table*}

\end{document}